\documentclass[12pt,preprint]{aastex}
\newcommand{\beq}{\begin{equation}}
\newcommand{\beqa}{\begin{eqnarray}}
\newcommand{\eeq}{\end{equation}}
\newcommand{\eeqa}{\end{eqnarray}}
   
\newcommand{\etal}{{\it et al. }}

\newcommand{\lsim}{\la}
\newcommand{\gsim}{\ga}

\newcommand{\bfxi}{\mbox{\boldmath{$\xi$}}}
\newcommand{\bfeta}{\mbox{\boldmath{$\eta$}}}

\shorttitle{}
\shortauthors{Takahashi \& Nakamura}

\begin{document}

\title{
Wave Effects in Gravitational Lensing of Gravitational Waves 
from Chirping Binaries
}
\author{Ryuichi Takahashi and Takashi Nakamura}
\affil{
Department of Physics, Kyoto University,
Kyoto 606-8502, Japan 
}


\begin{abstract}
In the gravitational lensing of gravitational waves, 
  the wave optics should be used 
 instead of the geometrical optics when the
 wavelength $\lambda$ of the gravitational waves is longer than 
 the Schwarzschild radius of the lens mass $M_L$.
For the gravitational lensing of the chirp signals from the coalescence of the super massive black holes at the redshift $z_S\sim 1$ relevant to  LISA, 
the wave effects become important for the lens mass smaller than $\sim 10^8 M_{\odot}$. For such cases, we compute how accurately we can extract
 the mass of the lens and the source position from the lensed signal.
We consider two simple lens models:
 the point mass lens and the SIS (Singular Isothermal Sphere).
We find that the lens mass and the source
 position can be determined within $\sim 0.1 \%~[(S/N)/10^3]^{-1}$ 
 for the lens mass larger than $10^8 M_{\odot}$
 and $\gsim 10 \%~[(S/N)/10^3]^{-1}$ for the lens mass smaller than $10^7 M_{\odot}$ due to the diffraction effect, where $(S/N)$ is the signal to noise
ratio of the unlensed chirp signals.
For the SIS model, if the source position is outside the Einstein radius,
only a single image exists in the geometrical optics approximation
so that the lens parameters can not be determined. While in the
wave optics cases  we find that  the lens mass  can be determined even 
for $M_L < 10^8 M_{\odot}$. For the point mass lens,
 one can extract the lens parameters even if the
 source position is far outside the Einstein radius.
As a result, the lensing cross section is an order of
 magnitude larger than that for the usual strong lensing of light. 
\end{abstract}



\section{Introduction}
Inspirals and mergers of compact binaries are the most promising
 gravitational wave sources and will be detected 
 by the ground based as well as the space based detectors in the near
 future (e.g. Cutler \& Thorne 2002).
Laser interferometers are now coming on-line or planned
 on broad frequency bands:
 for the high frequency band $10-10^4$ Hz, the ground based
 interferometers such as TAMA300, LIGO, VIRGO and GEO600 will be operated;
 for the low frequency band $10^{-4}-10^{-1}$ Hz, the Laser Interferometer
 Space Antenna\footnote{See http://lisa.jpl.nasa.gov/index.html}\ (LISA)
 will be in operation; for the intermediate frequency band $10^{-2}-1$ Hz,
 the space based interferometers such as DECIGO
 (Seto, Kawamura \& Nakamura 2001) are planned.
For the templates of the chirp signals from coalescing compact binaries,
 the post-Newtonian computations of the waveforms have been done by many
 authors. Using the matched filter techniques with the template,  
 we can obtain the binary parameters such as 
 the  mass  and the spatial position of the source 
(e.g. Cutler \& Flanagan 1994).

If the gravitational waves from coalescing binary 
 pass near massive objects,
 gravitational lensing should occur in the same way as it does for light.
The gravitational lensing of light is usually treated
 in the geometrical optics approximation, which is valid in all the
 observational situations
 (Schneider, Ehlers \& Falco 1992; Nakamura \& Deguchi 1999).
However for the gravitational lensing of gravitational waves,
 the wavelength is long so that the geometrical optics approximation is 
 not valid in some cases.    For example,      
 the wavelength $\lambda$ of the gravitational waves
 for the space interferometer is $\sim 1$ AU which is
 extremely larger than that of a visible light
 ($\lambda \sim 1 \mu$ m).
As shown by several authors (Ohanian 1974, Bliokh \& Minakov 1975,
 Bontz \& Haugan 1981, Thorne 1983, Deguchi \& Watson 1986a),
 if the wavelength $\lambda$ is larger than the Schwarzschild radius
 of the lens mass $M_L$, the diffraction effect is important
 and the magnification is small.
To see the reason why the ratio $M_L/\lambda$ determines the
 significance of the diffraction, we consider
 a double slit with the slit width comparable to 
 the Einstein radius $\xi_E \sim (M_L D)^{1/2}$ where
 $D$ is the distance
 from the screen
 to the slit (Nakamura 1998).
When waves with the wavelength $\lambda$ pass through the slit,
 the interference pattern is produced on the screen. 
The width ${\ell}$ of the central peak of the interference pattern
 is $\ell \sim (D/\xi_E) \lambda$.
Then the maximum magnification of the wave flux is of the order
 $\sim \xi_E/\ell \sim M_L/\lambda$.    
Thus the diffraction effect is important for
\beq
  M_L \lsim 10^8 M_{\odot} \left( \frac{f}{{\mbox{mHz}}} \right)^{-1},
\eeq
 where $f$ is the frequency of the gravitational waves.
However as suggested by Ruffa (1999),
 the focused region by the gravitational lensing
 would have a relatively large area because of the diffraction, so that 
 the lensing probability will increase.
Since the gravitational waves from 
 the compact binaries are coherent, the interference
 is also important (Mandzhos 1981, Ohanian 1983,
 Schneider \& Schmid-Burgk 1985, Deguchi \& Watson 1986b,
 Peterson \& Falk 1991).
Thus we expect that the wave effects (diffraction and interference)
 would provide much information about the lens objects.

In this paper, we consider the wave effects in the gravitational lensing
 of gravitational waves.
We take the coalescence of the super massive black holes (SMBHs) of
 mass $10^4-10^7 M_{\odot}$ as the sources. 
SMBH binary is one of the most promising sources 
 for LISA and will be detected with very high signal to noise ratio,
 $S/N \sim 10^3$ (Bender \etal 2000).
Since the merging SMBHs events will be detected
 for extremely high redshift ($z > 5$),
 the lensing probability is relatively high
 and hence some lensing events are expected.
We consider the two simple lens models; 1) the point mass lens in
 which compact objects (such as black holes) are assumed as lens
 objects and 2) the SIS (Singular Isothermal Sphere) lens in which
 galaxies, star clusters and CDM (Cold Dark Matter) halos are
 assumed as lens. 
The wave effects become important for the lens mass $10^6-10^9
 M_{\odot}$ which is determined by
 the LISA band, $10^{-4}$ to $10^{-1}$ Hz from Eq.(1).
The frequency of the gravitational waves from the coalescing
 SMBH binary chirps so that we could see wave effects for different
 frequency in the lensed chirp signals.

We calculate the gravitational lensed waveform using the wave optics 
  for the two lens models: the point mass lens and the SIS.
Then, we investigate how accurately we can extract the information on the lens
 object from the gravitational lensed signals detected by LISA 
 using the Fisher-matrix formalism (e.g. Cutler \& Flanagan 1994).
Cutler (1998) studied the estimation errors for the merging
 SMBHs by LISA (see also Vecchio \& Cutler 1998; Hughes 2002;
 Moore \& Hellings 2002; Hellings \& Moore 2002; Seto 2002; Vecchio 2003). 
Following Cutler (1998),
 we calculate the estimation errors,
 especially for the lens mass and the source position.
We assume the 1 yr observation before the final merging 
 and consider the lens mass in the range $10^6-10^9 M_{\odot}$. 
Then the typical time delay between the double images is $10-10^4$ sec
 which is much smaller than 1 yr.  

This paper is organized as follows.
In \S 2, we briefly review the wave optics in gravitational lensing
 for the point mass lens and the SIS model.  
In \S 3, we discuss the gravitational lensed waveforms detected
 with LISA, and mention the parameter estimation based
 on the matched filtering analysis.  
In \S 4, we numerically evaluate the signal to noise ratio and 
 the parameter estimation errors.
We discuss the dependence of the estimation errors on the lens model, 
 the lens mass and the source position.
In \S 5, we estimate the lensing event rate. 
\S 6 is devoted to summary and discussions.
We assume the $(\Omega_M,\Omega_{\Lambda})=(0.3,0.7)$ cosmology and 
the Hubble parameter $H_0=70$ km/sec/Mpc, and use the units of $c=G=1$.

\section{Wave Optics in Gravitational Lensing}

In this section, we briefly review the wave optics in the gravitational lensing
 of the gravitational waves (Schneider, Ehlers \& Falco 1992;
 Nakamura \& Deguchi 1999).
We consider the gravitational waves propagating under the gravitational
 potential of the lens object.
The  metric is given by
\beq
  ds^2 = - \left( 1+2U \right) dt^2 + \left( 1-2U \right)
 d \mathbf{r}^2 \equiv g_{\mu\nu}^{({\rm B})}dx^\mu dx^\nu,
\label{metric}  
\eeq 
where $U(\mathbf{r})~(\ll 1)$ is the gravitational potential of
 the lens object. Let us consider the linear perturbation $h_{\mu\nu}$ in the
background metric tensor $ g_{\mu\nu}^{({\rm B})}$ as
\beq
g_{\mu\nu}=g_{\mu\nu}^{({\rm B})}+h_{\mu\nu}.
\eeq
Under the transverse traceless Lorentz gauge condition of
 $h^\nu_{\mu;\nu}=0$ and ${h^\mu}_{\mu}=0$ we have 
\beq
 {h_{\mu\nu ;\alpha}}^{;\alpha}+2R^{({\rm B})}_{\alpha\mu\beta\nu}
 h^{\alpha\beta} =0,
\eeq
where $;$ is the covariant derivative with respect 
to $ g_{\mu\nu}^{({\rm B})}$ and  $R^{({\rm B})}_{\alpha\mu\beta\nu}$
is the background Riemann tensor. 
If the wavelength $\lambda$ is much smaller than the typical radius of
the curvature of the background, we have
\beq
 {h_{\mu\nu ;\alpha}}^{;\alpha}=0.
\eeq

Following the eikonal approximation to the above equation by Baraldo, Hosoya
and Nakamura (1999), we express the gravitational wave as
\beq
h_{\mu \nu}= \phi~e_{\mu \nu},
\eeq
where  $e_{\mu \nu}$ is the  the polarization tensor of the
 gravitational wave (${e^\mu}_{\mu}=0, e_{\mu \nu} e^{\mu \nu}=2$)
 and $\phi$ is a scalar.
 The polarization tensor $e_{\mu \nu}$ is parallelly transported along
 the null geodesic ($e_{\mu \nu ;\alpha} k^\alpha=0$, where $k^\alpha$ is
 a wave vector) (Misner, Thorne \& Wheeler 1973).
Then the change of the polarization tensor by gravitational lensing
 is of the order of $U~(\ll 1)$
 which is very small in our observational situation, and hence
 we can regard the polarization tensor as a constant. 
Thus, we treat the scalar wave $\phi$,  
 instead of the gravitational wave $h_{\mu \nu}$,
 propagating through the curved space-time.
The propagation equation of the scalar wave is  
\beq
 \partial_{\mu} (\sqrt{-g^{({\rm B})}} g^{({\rm B}) \mu \nu}
 \partial_{\nu} \phi)=0.
\label{prop}
\eeq
For the scalar wave in the frequency domain $\tilde{\phi}(f,\mathbf{r})$,
 the above equation (\ref{prop}) with Eq.(\ref{metric}) is
 rewritten as, 
\beq
   \left( \nabla^2 + \omega^2 \right) \tilde{\phi} = 4 \omega^2 U
 \tilde{\phi},
\label{propf}
\eeq
where $\omega=2 \pi f$.
The above equation (\ref{propf}) can be solved by using the 
 Kirchhoff integral theorem (Schneider, Ehlers \& Falco 1992).

It is convenient to define the amplification factor as 
\beq
F(f)=\tilde{\phi}^L(f) / \tilde{\phi}(f),
\eeq
where $\tilde{\phi}^L(f)$  and
 $\tilde{\phi}(f)$ are the lensed and unlensed ($U=0$ in
 Eq.(\ref{propf})) gravitational wave amplitudes, respectively.
In Fig.1, we show the gravitational lens geometry of the source, the lens and
the  observer.
$D_L, D_S$ and $D_{LS}$ are the distances 
 to the lens, the source and from the source to the lens, respectively.
$\bfeta$ is a position vector of the source in the source plane
while  $\bfxi$ is the impact parameter in the lens plane.
We use the thin lens approximation in which the lens is characterized 
 by the surface mass density $\Sigma(\bfxi)$ and the gravitational waves
 are scattered on the thin lens plane. 
Then, the amplification factor $F(f)$ at the observer is
 given by (Schneider, Ehlers \& Falco 1992),
\beq
F(f)=\frac{D_S ~\xi_0^2}{D_L D_{LS}} \frac{f}{i}
 \int d^2 \mathbf{x} ~\exp\left[~2 \pi i f t_d(\mathbf{x},\mathbf{y}) \right],
\label{ampfm} 
\eeq
where $\mathbf{x}=\bfxi/\xi_0$, and $\mathbf{y}=\bfeta D_L/\xi_0 D_S$
 is the source position.
$\xi_0$ is the arbitrary normalization constant of the length.
$t_d$ is the arrival time at the observer from the source.
 $F$ is normalized such that $|F|=1$ in no lens limit ($U=0$).

Though we do not take account of the cosmological expansion in the
 metric, Eq.(\ref{metric}), we can  apply  the result without  
the cosmological expansion to cosmological situations  
since the wavelength of the gravitational
waves is much smaller than the horizon scale. What we should do 
  is 1) take the angular diameter distances
 and 2) replace $f$ with $f (1+z_L)$ where $z_L$ is the redshift of the
 lens (Baraldo, Hosoya and Nakamura 1999).
Then the amplification factor $F(f)$ in Eq.(\ref{ampfm}) is
 rewritten in cosmological situations as,
\beq
 F(f)=\frac{D_S ~\xi_0^2 (1+z_L)} {D_L D_{LS}} \frac{f}{i}
 \int d^2 \mathbf{x} ~\exp\left[~2 \pi i f t_d(\mathbf{x},\mathbf{y})
 \right],
 \label{ampf} 
\eeq
where $D_L$, $D_S$ and $D_{LS}$ denote the angular diameter distances. 
The arrival time $t_d$ at the observer from the source position
 $\bfeta$ through $\bfxi$  is  given by 
 (Schneider, Ehlers \& Falco 1992),
\beq
 t_d(\mathbf{x},\mathbf{y})=\frac{D_S \xi_0^2}{D_L D_{LS}} \left( 1+z_L \right)
 \left[ \frac{1}{2} \left| \mathbf{x-y} \right|^2 - \psi(\mathbf{x}) 
  + \phi_m (\mathbf{y}) \right],
\eeq
The nondimensional deflection potential $\psi(\mathbf{x})$ is determined by 
\beq
\nabla_x^2 \psi = 2 \Sigma / \Sigma_{cr},
\eeq
 where $\nabla_x^2$ denotes the two-dimensional Laplacian with respect
 to $\mathbf{x}$,  $\Sigma$  is the surface mass density of the lens and $\Sigma_{cr}=D_S/(4 \pi D_L D_{LS})$.
We choose $\phi_m (\mathbf{y})$ so that the minimum value of
 the arrival time is zero.
We derive $\psi(\mathbf{x})$ and $\phi_m (\mathbf{y})$ in the following
 subsections for the point mass lens and the SIS model.

In the geometrical optics limit ($f \gg t_d^{-1}$), the stationary 
 points of the $t_d(\mathbf{x},\mathbf{y})$ contribute to the integral of
 Eq.(\ref{ampf}) so that the image positions
 $\mathbf{x}_j$ are determined by the lens equation
\beq
 \partial t_d(\mathbf{x},\mathbf{y})/\partial \mathbf{x} =0. 
\eeq
 This is just the Fermat's principle.   
The integral on the lens plane
 (in Eq.(\ref{ampf})) is reduced to the sum over these
 images (Nakamura \& Deguchi 1999) as
\beq
 F(f)=\sum_j \left| \mu_j \right|^{1/2} \exp
 \left[~2 \pi i f t_{d,j}- i \pi n_j \right],
\label{gf}
\eeq
where the magnification of the j-th image is $\mu_j=1/\det \left( 
 \partial \mathbf{y} / \partial \mathbf{x}_j \right)$, 
 $t_{d,j}=t_d (\mathbf{x}_j,\mathbf{y})$ and $n_j=0,1/2,1$ when
 $\mathbf{x}_j$ is a minimum, saddle, maximum point of
 $t_d (\mathbf{x},\mathbf{y})$.
In the time domain the wave is expressed as
\beq
\phi^L(t,{\bf r})=\sum_j \left| \mu_j \right|^{1/2}\phi(t-t_{d,j},{\bf r} )
 \exp \left[ -i \pi n_j \right].
\eeq
This shows that the oscillatory behavior of  $F(f)$ in high
 frequency $f$ is essential to obtain the time delay among the images.

\subsection{Point Mass Lens}
The surface mass density
 is expressed as  $\Sigma(\bfxi)=M_L \delta^2(\bfxi)$ 
where  $M_L$ is the lens mass.  
As the normalization constant $\xi_0$ we adopt the Einstein radius
 given by  $\xi_0=(4 M_L D_L D_{LS}/D_S)^{1/2}$ while
 the nondimensional deflection potential is $\psi(\mathbf{x})=\ln x$.
In this case, Eq.(\ref{ampf}) is analytically integrated as (Peters 1974),
\beqa
  F(f) &=& \exp\left[ \frac{\pi w}{4} + i\frac{w}{2}
 \left( \ln \left(  \frac{w}{2} \right)
 -2 \phi_m(y) \right)\right]  \nonumber \\
 &&~~~~~\times \Gamma \left( 1- \frac{i}{2} w \right) ~_1 F_1 \left(
 \frac{i}{2} w,1;\frac{i}{2} wy^2 \right),
\eeqa
where $w=8 \pi M_{L z} f$; $\phi_m(y)=(x_m-y)^2/2-\ln{x_m}$ with
 $x_m=(y+\sqrt{y^2+4})/2$; $M_{L z}=M_L (1+z_L)$ is the redshifted 
 lens mass and $_1 F_1$ is the confluent hypergeometric function.
Thus, the amplification factor $F(f)$ includes the two lens
 parameters; the redshifted lens mass $M_{L z}$
 and the source position $y$. 
In the geometrical optics limit $(f \gg M_{L z}^{-1})$ from Eq.(\ref{gf})
 we have 
\beq
  F(f) = \left| \mu_{+} \right|^{1/2}-i \left| \mu_{-}
   \right|^{1/2} e^{2 \pi i f \Delta t_d},
\label{gpf}  
\eeq
where the magnification of each image is $\mu_{\pm}=1/2 \pm (y^2+2)
 /(2 y \sqrt{y^2+4})$ and the time delay between the double images is 
 $\Delta t_d = 4 M_{L z} [ y \sqrt{y^2+4}/2 
 + \ln ((\sqrt{y^2+4}+y)/(\sqrt{y^2+4}-y))]$. 
The typical time delay is $\Delta t_d \sim 4 M_{L z} = 2 \times 10^3
 ~\mbox{sec} ~(M_{L z}/10^8 M_{\odot})$.

\subsection{Singular Isothermal Sphere}
The surface density of the SIS (Singular Isothermal Sphere) is characterized
 by the velocity dispersion $v$ as, $\Sigma(\bfxi)=v^2/(2 \xi)$. 
As the normalization constant we adopt the Einstein radius
  $\xi_0=4 \pi v^2 D_L D_{LS}/D_S$ and
 the nondimensional deflection potential is $\psi(\mathbf{x})=x$.
In this case $F(f)$ in Eq.(\ref{ampf}) is expressed as
\beq
  F(f)=-i w e^{i w y^2/2} \int_0^{\infty} dx ~x~ J_0 (w x y)
 \exp \left[ i w \left( \frac{1}{2} x^2 - x + \phi_m (y) \right) \right],
\eeq 
where $J_0$ is the Bessel function of zeroth order; $\phi_m(y)=y+1/2$
 and $w=8 \pi M_{L z} f$ where $M_{L z}$ is defined as the mass inside
 the Einstein radius given by 
 $M_{L z}=4 \pi^2 v^4 (1+z_L) D_L D_{LS}/D_S$.
Then, $F(f)$ depends on the two lens parameters $M_{L z}$ and $y$.
We computed the above integral numerically for various parameters.
In the geometrical optics limit $(f \gg M_{L z}^{-1})$, 
 $F$ is given by,
\beqa
  F(f) &=& \left| \mu_{+} \right|^{1/2}-i \left| \mu_{-} \right|^{1/2}
     e^{2 \pi i f \Delta t_d}  ~~~~~\mbox{for}~y \leq 1,  \nonumber \\
       &=& \left| \mu_{+} \right|^{1/2}  \hspace{3.95cm}
       \mbox{for}~y \geq 1,    
\label{gsf}
\eeqa
where $\mu_{\pm}=\pm 1 +1/y$ and $\Delta t_d= 8 M_{L z} y$. 
If $y \leq 1$, double images are formed.

The wave effects in gravitational lensing of gravitational waves were
 discussed for the point mass lens (Nakamura 1998;
 Ruffa 1999; De Paolis \etal 2002; Zakharov \& Baryshev 2002) and a
 Kerr BH (Baraldo, Hosoya \& Nakamura 1999).
However as far as we know the wave effects for the SIS (Singular Isothermal
 Sphere) model have not been discussed although  
the SIS model can be used for more realistic lens objects such as 
 galaxies and star clusters.  

\subsection{The Amplification Factor}
In Fig.2, we show the amplification factor $|F(f)|$ as a function of
 $w~(=8 \pi M_{L z} f)$ for the fixed source position $y=0.1,0.3,1,3$ 
 for the point mass lens (left panel) and the SIS lens (right panel).
For $w \lsim 1$, the amplification is very small due to the diffraction
 effect (e.g., Bontz \& Haugan 1980). Since in this case the wave length is
so long that the wave does not feel the existence of the lens.
For $w \gsim 1$,  $|F(f)|$ asymptotically converges to the geometrical 
 optics limit (Eq.(\ref{gpf}) and (\ref{gsf}));
\beq
 \left| F(f) \right|^2 = \left| \mu_{+} \right| + \left| \mu_{-} \right| 
  + 2 \left| \mu_{+} \mu_{-} \right|^{1/2} \sin ( 2 \pi f \Delta t_d ),
\label{absf}
\eeq
where  $\mu_{-}=0$ for $y \geq 1$ in the SIS.
The first and second terms in Eq.(\ref{absf}), $ | \mu | =
 | \mu_{+} | + | \mu_{-} |$, represent the total magnification
 in the geometrical optics. 
The third term expresses the interference between the double images.
The oscillatory behavior (in Fig.2) is due to this interference.
The amplitude and the period of this oscillation are approximately
 equal to 
 $2 |\mu_+ \mu_-|^{1/2}$ and  $2 \pi f \Delta t_d$ in the third 
 term of Eq.(\ref{absf}), respectively.
As the source position $y$ increases, the total magnification $|\mu|$ 
 $(=|\mu_+|+|\mu_-|)$ and 
 the amplitude of the oscillation $2 |\mu_+ \mu_-|^{1/2}$ decrease.
This is because each magnification $|\mu_{\pm}(y)|$ decreases as 
 $y$ increases.
We note  that even for $y \ge 1$  in SIS model ($y=3$ in Fig.2)
   the damped oscillatory behavior appears,  
which looks like the time delay factor of $ \sin ( 2
 \pi f \Delta t_d )$ although only a single image exists in the
 geometrical optics limit.

Fig.3 is the same as Fig.2, but we show the phase of the amplification
 factor $\theta_F(f)=-i \ln [F(f)/|F(f)|]$.  
The behavior is similar to that of the amplitude (in Fig.2),
 and the wave effects appear in the phase $\theta_F$ 
 as well as the amplitude $|F|$.  
For $w \gsim 1$,  $\theta_F(f)$ asymptotically converges to the geometrical 
 optics limit (Eq.(\ref{gpf}) and (\ref{gsf}));
\beq
 \theta_F(f)=\arctan \left[ \frac{- |\mu_-|^{1/2} \cos (2 \pi f \Delta
 t_d)}{|\mu_+|^{1/2} + |\mu_-|^{1/2} \sin (2 \pi f \Delta t_d)} \right],
\label{absfp}
\eeq
where $\mu_{-}=0$ for $y \geq 1$ in the SIS.
From the above equation (\ref{absfp}),
the phase $\theta_F$ oscillates between $-\arctan(|\mu_-/\mu_+|^{1/2})$ 
 and $\arctan(|\mu_-/\mu_+|^{1/2})$ with the period of 
 $2 \pi f \Delta t_d$.
As the source position $y$ increases, the magnification ratio
 $|\mu_-/\mu_+|$ and 
 the amplitude of the oscillation $\arctan(|\mu_-/\mu_+|^{1/2})$
 decrease.

\section{Gravitational Lensed Waveform and Parameter Estimation}

\subsection{Gravitational Wave Measurement with LISA}
We briefly discuss the gravitational wave measurement with LISA
 (see Cutler 1998; Bender \etal 2000).
LISA consists of three  spacecrafts forming an equilateral triangle 
 and orbits around the Sun, trailing $20^{\circ}$ behind the Earth. 
The sides of the triangle are $L=5 \times 10^6$ km in length, and the
 plane of the triangle is inclined at $60^{\circ}$ with respect to
 the ecliptic. The triangle rotates annually. 
The gravitational wave signal is reconstructed from the three data
 streams  that effectively correspond to three time-varying  
 armlength data.
Two of the three data are linearly independent of each other.
The data contain both gravitational waves signals
 to be fitted by matched filtering and noises
 which are assumed to be stationary, Gaussian
 and uncorrelated with each other (Cutler 1998).
The gravitational wave signals $h_{I,II}(t)$ 
 from a binary are written as 
\beq
  h_{I,II}(t)=\frac{\sqrt{3}}{2} \left[ F^{+}_{I,II}(t)h_+(t)
 +F^{\times}_{I,II}(t)h_{\times}(t) \right],
\label{ct}
\eeq
where $F^{+,\times}_{I,II}(t)$ are the pattern functions which depend 
 on the source's angular position of the binary,
 its orientation and detector's configuration. 
The quantities $h_{+,\times}(t)$ are the two polarization modes of 
 gravitational radiation  from  the  binary.  
The direction and the orientation
 of the binary and the direction of
 the lens are assumed to be constant during the observation
 in a fixed barycenter frame of the solar system.
Further discussion and details about the pattern functions are shown in
 Cutler (1998).

\subsection{Gravitational Lensed Signal Measured by LISA}

We consider the SMBH binaries at redshift $z_S$ as the sources.
We use restricted post-Newtonian approximation 
 as the in-spiral waveform \cite{cf94}.
The coalescing time for circular orbit is typically $t_c=0.1 \mbox{yr}~
 (\mathcal{M}_z/10^6 M_{\odot})^{-5/3} (f/10^{-4} \mbox{Hz})^{-8/3}$ where
 $\mathcal{M}_z=(M_1 M_2)^{3/5} (M_1 + M_2)^{-1/5} (1+z_S)$
 is the redshifted chirp mass.
At the solar system barycenter, the unlensed waveforms
 $\tilde{h}_{+,\times}(f)$ in the frequency domain are given by
\beqa
 \tilde{h}_{+}(f) &=& \mathcal{A} \left[ 1+ \left( \mathbf{L} \cdot
 \mathbf{n} \right)^2 \right] f^{-7/6} e^{i \Psi(f)}, \nonumber \\
 \tilde{h}_{\times}(f) &=& -2 i \mathcal{A} \left( \mathbf{L} \cdot
 \mathbf{n} \right) f^{-7/6} e^{i \Psi(f)}, 
\label{uhc}
\eeqa
where $\mathbf{L}$ (given by $\bar{\theta}_L,\bar{\phi}_L$) is the unit
 vector in the direction of the binary's orbital angular
 momentum and $\mathbf{n}$ (given by $\bar{\theta}_S,\bar{\phi}_S$) is the
 unit vector toward the binary. 
These vectors are defined in a fixed barycenter frame of the solar system.
The amplitude $\mathcal{A}$ and the phase $\Psi(f)$ depend on six
 parameters; the redshifted chirp mass $\mathcal{M}_z$ and reduced mass
 $\mu_z=M_1 M_2 (1+z_S)/(M_1+M_2)$; the spin-orbit coupling constant
 $\beta$; a coalescence time $t_c$ and phase $\phi_c$;
 the angular diameter distance to the source $D_S$.
The amplitude is
\beq
 \mathcal{A}= \sqrt{\frac{5}{96}} \frac{\pi^{-2/3}
 \mathcal{M}_z^{5/6}}{D_S (1+z_S)^2}.
\label{amp}
\eeq
where $D_S (1+z_S)^2$ is the luminosity distance to the source,
 and $\Psi(f)$ is a rather complicated function of $\mathcal{M}_z$,
 $\mu_z$, $\beta$,
 $\phi_c$ and $t_c$ (see Eq.(3.24) of Cutler \& Flanagan 1994).

The gravitational lensed waveforms $\tilde{h}^L_{+,\times}(f)$
 in the frequency domain are given by the product of the amplification
 factor $F(f)$ and the unlensed waveforms $\tilde{h}_{+,\times}(f)$
 (see section 2);
\beq
 \tilde{h}^L_{+,\times}(f)=F(f) ~\tilde{h}_{+,\times}(f).
\label{hh}
\eeq
where the function $F(f)$ is given in Eq.(\ref{ampf}).
Using   Eq.(\ref{ct}),(\ref{uhc}) and (\ref{hh}), the observed
 lensed signals $\tilde{h}^L_{\alpha}(f)$ ($\alpha=I,II$)
 with LISA are given in the stationary phase approximation as,
\beqa
 \tilde{h}^L_{\alpha}(f) &=& \frac{\sqrt{3}}{2}
 \frac{D_S ~\xi_0^2 \left( 1+z_L \right)}{D_L D_{LS}} \frac{f}{i} 
 \int d^2 \mathbf{x} ~\Lambda_{\alpha}(t+t_d(\mathbf{x},\mathbf{y})) 
 e^{2\pi i f t_d(\mathbf{x},\mathbf{y})} e^{-i \left( \phi_D + \phi_{p,\alpha} 
 \right) (t+t_d(\mathbf{x},\mathbf{y}))} \nonumber \\
 &&~~~~~~\times \mathcal{A} f^{-7/6} e^{i \Psi(f)}, 
\label{lensig}
\eeqa       
where $\phi_{p,\alpha}(t)=\tan^{-1} [ 2 (\mathbf{L} \cdot \mathbf{n})
 F^{\times}_{\alpha}(t)/ \{ 1+ (\mathbf{L} \cdot \mathbf{n})^2 \}
 F^{+}_{\alpha}(t) ]$ and $\Lambda_{\alpha}(t)=[ ( 2 
 ~\mathbf{L} \cdot \mathbf{n})^2 F^{\times~2}_{\alpha}(t)+ 
 \{ 1+ (\mathbf{L} \cdot \mathbf{n})^2 \}^2 F^{+~2}_{\alpha}(t) ]^{1/2}$.
The Doppler phase is $\phi_D(t)=2 \pi f(t) R \sin \bar{\theta}_S
 \cos \left( \bar{\phi}(t) - \bar{\phi}_S \right)$,
 $R=1$ AU and $\bar{\phi}(t) =2 \pi t/T~(T=1$ yr). 
$t=t(f)$ is given in Eq.(3.10) of Cutler \& Flanagan (1994).
In no lens limit of $\psi(\mathbf{x})=0$, the lensed signals
 $\tilde{h}^L_{\alpha}(f)$ in Eq.(\ref{lensig}) agree with
 the unlensed ones $\tilde{h}_{\alpha}(f)$ in Cutler (1998).
We assume the source position $\mathbf{y}$ is constant during the
 observation, since the characteristic scale of the interference
 pattern, $\sim 10^7 {\mbox{AU}} (M_{L z}/10^8 M_{\odot})^{-1/2}
 (f/{\mbox{mHz}})^{-1}$ $[(D_S D_L/D_{LS})/{\mbox{Gpc}}]^{1/2}$,
 is extremely larger than the LISA's orbital radius ($1$ AU).

Since the lensed signals $\tilde{h}^L_{\alpha}(f)$
 in Eq.(\ref{lensig}) are given by double integral, we approximate
 $\tilde{h}^L_{\alpha}(f)$ in the two limiting cases; 
 1) geometrical optics limit $(f \gg t^{-1}_{d})$ and 2) the  
 time delay being much smaller than LISA's orbital period of 
 $(t_{d} \ll 1$ yr).
In the geometrical optics limit, from Eq.(\ref{gf}) we obtain,
\beq
  \tilde{h}^L_{\alpha}(f)=\frac{\sqrt{3}}{2} \sum_j 
 \left| \mu_j \right|^{1/2} 
 \Lambda_{\alpha}(t+t_{d,j}) 
 e^{2\pi i f t_{d,j}-i\pi n_j} e^{-i \left( \phi_D 
 + \phi_{p,\alpha} \right) (t+t_{d,j})} \times
 \mathcal{A} f^{-7/6} e^{i \Psi(f)} .
\eeq  
If the time delay is much smaller than LISA's orbital period
 $(t_{d} \ll 1$ yr), we expand $\Lambda_{\alpha},\phi_D$ and
 $\phi_{p,\alpha}$ around  $t_d=0$ as,
\beqa
 \tilde{h}^L_{\alpha}(f) &=& \frac{\sqrt{3}}{2}
 \Lambda_{\alpha}(t) e^{-i \left( \phi_D + \phi_{p,\alpha} \right) (t)} 
 \times \mathcal{A} f^{-7/6} e^{i \Psi(f)}  \nonumber \\
 && \times \left[~F(f) + \frac{d}{dt} \left\{ \ln \Lambda_{\alpha}
 -i \left( \phi_D + \phi_{p,\alpha} \right) \right\} 
 \frac{f}{2\pi i} \frac{d}{df} \left( \frac{F(f)}{f} \right)
 +\mathcal{O} \left( (t_d/1 \mbox{yr})^2 \right) \right].
\label{hca2}
\eeqa
Since we consider the lens mass $M_{L z}=10^6-10^9 M_{\odot}$,
 the time delay is much smaller than  1 yr.
Thus we use the above equation (\ref{hca2}) as the lensed
 waveforms for the following calculations.

In Fig.4, the lensed signals $| \tilde{h}^L_{\alpha}(f)|$ $(\alpha=I,II)$
 and the unlensed ones $| \tilde{h}_{\alpha}(f)|$ are shown.
We show the results from one year before the final merging to the inner most
 stable circular orbit (the binary separation is $r=6 (M_1+M_2)$).
We set typical parameters at the SMBH binary masses
 $M_{1,2 z}=10^6 M_{\odot}$, the lens mass
 $M_{L z}=10^8 M_{\odot}$ and the source position
 $y=1$ for the point mass lens.
The angular parameters are $\cos \bar{\theta}_S=0.3, \bar{\phi}_S=5.0,
 \cos \bar{\theta}_L=0.8, \bar{\phi}_L=2.0$, and the source redshift is
 $z_S=1$ (the angular diameter distance is $H_0 D_S=0.386$).
Therefore the frequency range is from $5 \times 10^{-5}$ to 
 $2 \times 10^{-3}$ Hz and the time delay is $4 \times 10^3$ sec. 
The strange behavior in the lower frequency $f \lsim 10^{-4}$ Hz is due to
 the LISA's orbital motion.  In this frequency region, however,
the difference between the lensed signal
 and the unlensed one is small due to the diffraction (see Fig.2).
On the other hand, 
 the oscillatory behavior appears in the higher frequency region 
 $f \gsim 10^{-4}$ Hz. This critical frequency is 
 determined by the inverse of the lens mass $8 \pi M_{L z}$ (see Fig.2).
The oscillatory amplitude and the period are determined by the product of
 the magnifications $2 |\mu_+\mu_-|^{1/2}=2/(y \sqrt{y^2+4})$ and the
 inverse of the time delay, $1/\Delta t_d$
 (see the third term of Eq.({\ref{absf}})).

\subsection{Parameter Extraction}

We briefly mention the matched filtering analysis and the parameter
 estimation errors \cite{finn92,cf94}. 
We assume that the signal $\tilde{h}^L_{\alpha}(f)$ is characterized by
 some unknown parameters $\gamma_i$.
In the present case, there are ten source parameters 
 ($\mathcal{M}_z$, $\mu_z$, $\beta$, $\phi_c$, $t_c$, $D_S$,
 ${\bar \theta_S}$, ${\bar \phi_S}$, ${\bar \theta_L}$,
 ${\bar \phi_L}$) and two lens parameters ($M_{L z},y$).
In the matched filtering analysis the variance-covariance matrix of the
parameter estimation error $\Delta \gamma_i$ is given by inverse of the
 Fisher information matrix $\Gamma_{ij}$ as $\langle \Delta  
 \gamma_i \Delta \gamma_j \rangle = \left( \Gamma^{-1} \right)_{ij}$.
The Fisher matrix becomes
\beq 
  \Gamma _{ij} 
 = 4 \sum_{\alpha=I,II} {\mbox{Re}} \int \frac{df}{Sn(f)}~
 \frac{\partial \tilde{h}_{\alpha}^{L *}(f)}{\partial \gamma_i}
 \frac{\partial \tilde{h}_{\alpha}^L(f)}{\partial \gamma_j},
\label{fis}
\eeq
where $Sn(f)$ is the noise spectrum.
The noise spectrum $Sn(f)$ is the sum of the instrumental
 and the confusion noise, and we adopt the same noise spectrum
 as that in Cutler (1998).
The signal to noise ratio ($S/N$) is given by
\beq
  (S/N)^{~2} = 4 \sum_{\alpha=I,II} \int \frac{df}{Sn(f)}~
 \left| \tilde{h}^L_{\alpha}(f) \right|^2.
\label{snr}
\eeq
We computed the variance-covariance matrix $\Gamma_{ij}$ for a wide
 range of the lens parameters ($M_{Lz},y$), using the lensed waveform
 in Eq.(\ref{hca2}).
Since the $S/N$ is very high for the SMBH merger, the Fisher matrix
 approach to calculate the estimation errors is valid (Cutler 1998).
We integrate gravitational lensed waveform (in Eq.(\ref{fis}) and
 Eq.(\ref{snr})) from 1 yr before the final merging to the cut-off frequency
 $f_{\mbox{cut}}$ when the binary separation becomes $r=6 (M_1 + M_2)$. 
We do not consider the low-frequency cut-off of LISA,
 which is low frequency noise wall of space-based instruments 
 and is around $10^{-5}-10^{-4}$ Hz (Vecchio 2003). 
This assumption is to underestimate the errors in
 estimation parameters.

\section{Results}
In this section, we present numerical results to compute the signal to 
 noise ratio (S/N) and the errors in estimation parameters. 
We randomly distribute 100 binaries over various directions and orientations  
 on celestial spheres at $z_S=1$ (the distance is $H_0 D_S=0.386$).
We present the mean value averaged for 100 binaries.

\subsection{Lensing Effects on the Signal to Noise Ratio}

We demonstrate the gravitational lensing effect on the signal to
 noise ratio ($S/N$).
In Fig.5, the increasing factor of $S/N$ by the gravitational 
 lensing for the point mass lens is shown for the fixed source 
  position $y=0.1, 0.3, 1, 3$ as a function of the lens mass $M_{L z}$.
The vertical axis is the $S/N$ with the gravitational lensing divided by
 the unlensed $S/N$. 
Four panels are shown for the various SMBH binary  
 masses $M_{1,2 z}=10^4,10^5,10^6,10^7 M_{\odot}$.
We show the mean value averaged for 100 binaries, but the dispersion is
 negligibly small (less than 5 \%). 
For the lens mass smaller than $10^6 M_{\odot}$ the
 magnification is very small irrespective of the SMBH binary masses
 due to the diffraction effects.
In this case the Schwarzschild radius of the lens mass $M_{Lz}$ is
 smaller than the wavelength of gravitational waves $\lambda \sim 1$ AU,
 and the waves are not magnified by lensing.
This critical lens mass ($10^6 M_{\odot}$) is mainly determined by
 the inverse of the knee frequency of the LISA's noise spectrum,
 $1/(8 \pi f) \sim 8 \times 10^6 M_{\odot} (f/\mbox{mHz})^{-1}$ 
 (see Fig.2).
But for $10^7+10^7 M_{\odot}$, the SMBH binary coalescences at the
 lower frequency ($f \sim 10^{-4}$ Hz),
 thus the critical lens mass is shifted for larger mass ($10^7 M_{\odot}$)
 as shown in the right bottom panel of Fig.5. 
This tells us that if the lens mass is smaller than
 $10^{6} M_{\odot}$, the effect of the lens is very small.
If the lens mass is larger than $10^7 M_{\odot}$, 
 the damped oscillatory behavior appears due to the interference between
 the two images, and the S/N converges to the geometrical optics
 limit, $|\mu|^{1/2}=(y^2+2)^{1/2}/[y^{1/2} (y^2+4)^{1/4}]$, which 
 is independent of the lens mass.
As $y$ increases from $0.1$ (solid line) to $3$ (dashed line), the
 amplification decreases since the magnifications of the two images
 ($|\mu_{\pm}(y)|$) decrease as $y$ increases (see also Fig.2).

Fig.6 is the same as Fig.5, but for the SIS lens model.  
The behavior is very similar to that in the point mass lens.
For the lens mass larger than $10^7 M_{\odot}$, the S/N converges to
 the geometrical optics limit,
 $|\mu|^{1/2}=(2/y)^{1/2}$ for $y \leq 1$ and 
 $|\mu|^{1/2}=(1+1/y)^{1/2}$ for $y \geq 1$. 
As $y$ increases from $0.1$ (solid line) to $3$ (dashed line), the
 amplification decreases (see also the right panel of Fig.2).

\subsection{Parameter Estimation for the Lens Objects}
In this section, we show the parameter estimation for the lens objects.
We show the results for the SMBH binary with masses $10^6+10^6 M_{\odot}$,
 because we found $S/N$ is higher than the other binary masses
 ($M_{1,2 z}=10^4,10^5$ and $10^7 M_{\odot}$).
We distribute the 100 binaries over the various directions and the
 orientations at $z_S=1$, and the mean value of the S/N without lensing 
 is 2600 in these 100 binaries. 
We show the mean value of errors averaged for 100 binaries,
 for $M_{L z} \lsim 10^{7} M_{\odot}$ the dispersion is relatively
 large ($\lsim 40 \%$), but for $M_{L z} \gsim 10^{7} M_{\odot}$ 
 the results converge to that in the geometrical optics limit and 
 the dispersion is negligibly small.     

In Fig.7, the estimation errors for the redshifted lens mass
 $\Delta M_{L z}$ (left panel) and the source position $\Delta y$
 (right panel) are shown as a function of $M_{L z}$ with
 $y=0.1, 0.3, 1, 3$ for the point mass lens.
We use the units of $S/N=10^3$, and the results ($\Delta M_{L z}$,
 $\Delta y$) scale as $(S/N)^{-1}$.  
For $M_{L z} \lsim 10^{7} M_{\odot}$ the estimation errors are
 relatively large $\gsim 10 \%$, since the effect of
 lensing on the signals is very small due to the diffraction.
For $M_{L z} \gsim 10^8 M_{\odot}$ the geometrical optics approximation
 is valid, and the errors converge to a constant in Fig.7.
The redshifted lens mass and
 the source position can be determined  
 up to the accuracy of $ \sim 0.1 \%$, as shown in Fig.7.
The errors in the geometrical optics limit
 are well fitted by (see Appendix A),
\beqa
 \frac{\Delta M_{L z}}{M_{L z}} &=& \frac{1}{S/N} \times
 \frac{\sqrt{y (y^2+2)} (y^2+4)^{5/4}}{2 \tau},  \nonumber \\
 \frac{\Delta y}{y} &=& \frac{1}{S/N} \times
 \frac{\sqrt{y^2+2} (y^2+4)^{3/4}}{2 \sqrt{y}},
\label{dmdyp}
\eeqa
where $S/N$ is in the unlensed case, and
 $\tau=\Delta t_d / 4M_{L z}=y \sqrt{y^2+4}/2 
 + \ln ((\sqrt{y^2+4}+y)/(\sqrt{y^2+4}-y))$.
Thus, one could determine the lens parameters, 
 the redshifted lens mass and the source
 position, up to the accuracy of $\sim (S/N)^{-1}$.
The above equations (\ref{dmdyp}) are valid
 if the time delay $\Delta t_d$ is much smaller than
 the LISA's orbital period $1$ yr. 
If the time delay $\Delta t_d$ becomes comparable to $1$ yr, the
 LISA's orbital motion affects the results.

Fig.8 is the same as Fig.7, but as a function of $y$.
For $y \gsim 1$, the errors are convergent to the geometrical optics
 limit of Eq.(\ref{dmdyp}) irrespective of the lens mass.  
As $y$ increases, the time delay $t_d$ increases,
 and the geometrical optics limit ($f t_d \gg 1$) is valid.
We note that even for $y \gsim 10$ one can extract the lens
 information.
In the case of light, the observable is the lensed flux
 which is proportional to the magnifications, $\propto |\mu_{\pm}|$,
 but for gravitational waves the observable is the lensed amplitude which is
 proportional to the square root of the magnifications
 $\propto |\mu_{\pm}|^{1/2}$.
For example, let us consider the case where
the flux ratio of a brighter image to a fainter
 one is $100:1$. Then  the amplitude ratio is $10:1$ so that  
 the fainter image can be observed even if the source position is far
 from the Einstein radius in the case of gravitational waves. 
Denoting the largest source position for which one can extract the lens
 parameters as $y_{cr}$, we approximate the errors in Eq.(\ref{dmdyp})
 for the large $y$ limit; $\Delta \gamma/\gamma \simeq (S/N)^{-1} y^2$,
 where $\gamma=M_{L z},y$.
Then we obtain
\beq
  y_{cr} \simeq 10 \left( \frac{\Delta \gamma / \gamma}{0.1}
 \right)^{1/2} \left( \frac{S/N}{10^{3}} \right)^{1/2}.
\label{ycr} 
\eeq 
Thus the lensing cross section ($\propto y_{cr}^2$) increases an order
 of magnitude larger than 
 that for the usual strong lensing of light ($y_{cr}=1$)
 (e.g. Turner, Ostriker \& Gott 1984).

In Fig.9, the estimation errors for the SIS model are shown. 
For $M_{L z} \lsim 10^{7} M_{\odot}$
 the behavior is similar for the point mass lens.
But for $M_{L z} \gsim 10^{8} M_{\odot}$ the behavior strongly
 depends on $y$.
In the geometrical optics approximation,
 the errors are given by (Appendix A),
\beq
 \frac{\Delta M_{L z}}{M_{L z}} = \frac{\Delta y}{y} =
 \frac{1}{S/N} \times \sqrt{\frac{2 (1-y^2)}{y}} ~~~~{\mbox{for}}
 ~ y \leq 1. 
\label{dmdys}
\eeq
and the lens parameters are not determined for $y \geq 1$.
We note that even for $y=3$  the lens parameters can 
 be extracted for $M_{L z} \sim 10^{6} - 10^{8} M_{\odot}$
 due to the wave effects.  
For $y=0.1$ and $0.3$, the asymptotic behavior of errors
 are somewhat smaller than the results
 in Eq.(\ref{dmdys}), because the order of $1/f$ term
 in $F(f)$ (which is neglected in the geometrical optics
 approximation $f \to \infty$) affects the results.
For $y=1$, the errors decrease as lens mass increases as shown in Fig.9,
 because the errors converge to the results in the geometrical optics
 limit of Eq.(\ref{dmdys}) which vanish at $y=1$.
As a result, if $y > 1$, the errors asymptotically increase
 with the increase of the lens mass, but
 if $y < 1$, they asymptotically converge to constants. 

Fig.10 is the same as Fig.9, but as a function of $y$.
We note that even for larger $y \gsim 1$ we can extract the lens
 information.
Thus the lensing probability ($\propto y^2$) to determine
 the lens parameters increases 
 as compared with the results in geometrical optics limit
 for the lens objects in the mass
 range $10^6-10^8 M_{\odot}$.

\subsection{Lensing Effects on the Estimation Errors
 of the Binary Parameters}

We discuss the gravitational lensing effects on the 
 estimation errors of the SMBH binary parameters.
We study five binary parameters; the redshifted chirp mass
 $\mathcal{M}_{cz}$, the reduced mass $\mu_{z}$, the distance
 to the source $D_S$ and the angular resolution
 $(\bar{\theta}_S,\bar{\phi}_S)$.   
We find that the estimation errors of these
 parameters decrease because S/N increases by lensing
 (see Fig.5 and 6).
The error $\Delta \gamma$ is roughly proportional to the inverse of
 the $S/N$ as $\Delta \gamma \propto (S/N)^{-1}$
 (see Eq.(\ref{fis}) and (\ref{snr})).

\subsection{Results for Various SMBH Masses and Redshifts}

So far we presented the results for equal mass SMBH binaries with
 redshift $z_S=1$.
In this section, we comment the results for the case of various
 (unequal) SMBH masses $10^4-10^7 M_{\odot}$ and redshifts $z_S=1-10$.

The critical lens mass in which the wave effects become important
 $(10^6-10^8 M_{\odot})$ is mainly determined by the inverse of 
 the knee frequency of the LISA's noise spectrum,
 $\sim 8 \times 10^6 M_{\odot} ~(f/\mbox{mHz})^{-1}$,
 independent of the binary mass (see section 4.1).
But for the massive total mass binary $(M_{1 z}+M_{2 z}) \gsim 10^7
 M_{\odot}$, the binary coalescences at the lower frequency
 $(\sim 10^{-4} \mbox{Hz})$, thus the critical lens mass is shifted for
 larger mass $(10^7-10^9 M_{\odot})$.
For the larger lens mass $M_{L z} \gsim 10^8 M_{\odot}$, the results
 (the S/N increase and the estimation errors) converge to that in
 the geometrical optics limit
 irrespective of the binary mass.    
The estimation errors in Fig.7-10 are the case of $10^6-10^6 M_{\odot}$
 binary at redshift $z_S=1$ and are normalized to $S/N=10^3$ and
 simply scale as $(S/N)^{-1}$.
In order to translate the results in the various unequal SMBH binaries,
 we present the $S/N$ for
 binary masses $M_{1,2 z}=10^4-10^7 M_{\odot}$ with redshifts
 $z_S=1,3,5,10$ in Table.1.  
We assume 1 yr observation of in-spiral phase
 before final merging.
The results are the mean value of 100 binaries which are randomly 
 distributed at each redshift,
 and the dispersion is relatively large $\sim 50 \%$.
From Table.1, one could translate the results in Fig.7-10 
 into errors in real situations.

We also comment the results for the case of only $h_I$ data available,
 while we used the combination of $h_I$ and $h_{II}$ data (see section 3.1).
In this case, the $S/N$ increase in Fig.5 and 6 are not changed, 
 but the estimation errors are slightly larger $(\sim 30 \%)$
 than that in Fig.7-9 for $M_{L z} \lsim 10^7 M_{\odot}$ if the errors
 are normalized to $S/N=10^3$.
We note that the $S/N$ is $\sqrt{2}$ times smaller than that in the case
 of the two data available in Table.1.

\section{Lensing Event Rate}

We discuss the event rate of merging SMBHs and estimate
 the lensing probability and the lensing event rate.
The expected rate of merging SMBHs detected by LISA 
 is in the range $0.1-10^2$ events per year
 (Haehnelt 1994,1998).
Recently, Wyithe \& Loeb (2002) suggested that
 some hundreds detectable events per year could be expected,  
 considering the merger rate at
 exceedingly high redshift ($z > 5-10$).
Thus we take $\sim$ 300 events per year as the merging event rate. 

We consider the lens objects distributed over the universe and calculate
 the lensing probability for each lens model.
For the point mass lens, we take the compact objects
 ($10^6-10^9 M_{\odot}$) such as black holes as lens.    
Denoting the mass density parameter of compact objects as
 $\Omega_{\mbox{co}}$, the lensing probability for a source
 at redshift $z_S$ is (Schneider, Ehlers \& Falco 1992),
\beq
  P (z_S) = \frac{3}{2} \Omega_{\mbox{co}} y_{cr}^2 \int^{z_S}_{0} dz_L
 \frac{(1+z_L)^2}{H(z_L)/H_0} \frac{H_0 D_{L S}(z_L,z_S)~H_0 D_L(z_L)}
 {H_0 D_S(z_S)},
\label{lpp}
\eeq
 where $H(z)$ is the Hubble parameter at redshift $z$.
The cosmological abundance of the compact objects in the mass range 
 $10^6-10^9 M_{\odot}$ is limited by $\Omega_{\mbox{co}} \leq 0.01$ by
 the search for multiple images in radio sources
 (Wilkinson \etal 2001; see also Nemiroff \etal 2001).
In Table 2, we show the upper limit on the lensing probability 
 for the point mass lens.
Since we set $y_{cr}=10$ (Eq.(\ref{ycr})), the lensing probability is
 one hundred times larger than that normally assumed for the strong
 lensing of light ($y_{cr}=1$).  
As shown in Table 2,
 the upper limit of the lensing probability is very high (almost 1)
 and is typically $\sim (\Omega_{\mbox{co}}/10^{-2})$.
The lensing event rate is the product of the merging rate
 ($\sim 300$ per year) and the lensing probability, 
 so that the lensing events will be $1$ event per year
 if $\Omega_{\mbox{co}}=10^{-4}$.

For the SIS model we take CDM halos $(10^6-10^9 M_{\odot})$
 as the lens objects (e.g. Narayan \& White 1988).
The lensing probability is
\beq
  P (z_S) = \pi y_{cr}^2 \int^{z_S}_0 dz_L 
 \frac{(1+z_L)^2}{H(z_L)/H_0} \frac{H_0 D_{L S}(z_L,z_S)~H_0 D_L(z_L)}
 {H_0 D_S(z_S)}
 \int^{10^9 M_{\odot}}_{10^6 M_{\odot}} dM_{L} v N_v(v,z_L),
\eeq
 where $N_v$ is the comoving number density of the lens and is assumed
 to be given by the Press-Schechter velocity function
 (Press \& Schechter 1974) with $\sigma_8=1$.
In Table 2, we show the lensing probability for the SIS model.
We set $y_{cr}=3$ (see Fig.9) and hence the lensing probability is
 almost ten times larger than that for light ($y_{cr}=1$).
As shown in Table 2,
 the lensing probability is typically $\sim 10^{-4}-10^{-3}$.
The merger rate is $\sim 300$ events per year at high redshift ($z > 5$), 
 then the lensing events would be $1$ event per year.

We note that the results in Table.2 are for the case of the 
 $S/N=10^3$, and are somewhat overestimated for the binaries of
 $S/N < 10^3$ in Table.1. 
For example, the lensing probability is proportional to $(S/N)$
 from Eq.(\ref{ycr}) for the point mass lens, and
 it is appropriate to use $y_{cr}=1$ 
 for $S/N < 10^3$ in the SIS.
In the case of the high event rate ($\sim 300 \mbox{events/year}$),
 many fainter signals ($S/N \ll 10^3$) are expected and
 we note that the errors in Fig.7-10 are worse for these binaries.

Next, we discuss how we can identify the lensing signal.
If the lensing event occurs,           
 the amplitude and the arrival time of the
 gravitational waves are changed by lensing.
But the other features (such as binary mass) are not changed.
Thus, if the two signals have the same binary
 parameters (such as chirp mass) except for the amplitude   
 and the arrival time, that would be a signature of
 gravitational lensing in the geometrical optics limit.
More generally, oscillatory behavior in the waveform $|\tilde{h}^L(f)|$
 is a signature of gravitational lensing (see Fig.4).
However it will be difficult to identify the source and
 the lens objects in the sky,
 since the angular resolution of the LISA is $\sim 1$ deg
 (see Cutler 1998).
Furthermore, the gravitational wave amplitude is changed by the lensing
 magnification and hence one must assume the lens model
 in order to determine the distance to the source.
(Effect of lensing on measuring the distance is recently
 discussed in Holz \& Hughes (2002).) 
As one determines the distance to the source $D_S (z_S)$, 
 the redshift $z_S (D_S)$ could be determined
 if the cosmological parameters are well known
 (see Hughes 2002).

\section{Summary}

We have discussed the gravitational lensing of gravitational waves
 from chirping binaries, taking account of the wave effects in
 gravitational lensing.
The SMBH binary is taken as the source detected by LISA,  
 and the two simple lens models are considered: the point mass lens
 and the SIS model.
We calculate the lensing effects on the signal to noise ratio (S/N)
 and how accurately the information of the lens object, its mass, can
 be extracted from the lensed signal.
As expected, for the lens mass smaller than $10^8 M_{\odot}$, the
 wave effects are very important to calculate the $S/N$ and the
 errors in the estimation parameters.
It is found for the lens mass smaller than $10^6 M_{\odot}$ the signals are
 not magnified by lensing due to the diffraction effect.    
For the lens mass larger than $10^8 M_{\odot}$ the lens parameters can
 be determined within (very roughly)
 $\sim 0.1 \%~[(S/N)/10^3]^{-1}$.
We note that the lensing cross section to determine the lens parameters 
 is order of magnitude larger than that for light.  

In this paper, we calculate the case for LISA.
But similar analysis can be done for other detectors.
For the ground-based interferometers (TAMA300, LIGO, VIRGO, GEO600),
 neutron star binaries are taken as the sources and the lens mass
 for which the wave effects become important is
 $10-10^4 M_{\odot}$.
Similarly for the space-based interferometers such as DECIGO
 (Seto, Kawamura \& Nakamura 2001), the important lens mass becomes    
 $10^{5}-10^{7} M_{\odot}$.
Since mergers of neutron star binaries will be detected
 at least several per year for LIGO II
 (Phinney 1991; Kalogera \etal 2001) and $\sim 10^5$ per year 
 for DECIGO, the lensing events would also be expected
 for other detectors.

\acknowledgments
We would like to thank Naoki Seto and Takeshi Chiba for useful
 comments and discussions.
This work was supported in part by
Grant-in-Aid for Scientific Research
of the Japanese Ministry of Education, Culture, Sports, Science
and Technology,
No.14047212 (TN), and No.14204024 (TN).

\appendix

\section{Estimation errors in the geometrical optics limit}
To evaluate the estimation errors $\Delta M_{L z}, \Delta y$ in the
 geometrical optics limit, we consider the simple waveform;
\beq
   \tilde{h}^L(f)=\left( \left| \mu_{+} \right|^{1/2}
 -i \left| \mu_{-} \right|^{1/2} e^{2 \pi i f
 \Delta t_d} \right) \times \tilde{h}(f),
\eeq
where $\tilde{h} \propto\mathcal{A}$ is the unlenesd
 signal and $\Delta t_d \propto M_{L z}$, with three parameters
 $\gamma_i=(\ln M_{L z},y,\ln \mathcal{A})$.
Then, the Fisher matrix $\Gamma_{ij} (i,j=1,2,3)$ in Eq.(\ref{fis})
 can be analytically obtained as,
\beqa
  \Gamma_{11} &=& \left( 2 \pi \Delta t_d \right)^2 \left| \mu_- \right|
 ~( f \tilde{h} | f \tilde{h} ), \nonumber \\
  \Gamma_{12} &=& 4 \pi^2 \Delta t_d
 \frac{\partial \Delta t_d}{\partial y} \left| \mu_- \right|
 ~( f \tilde{h} | f \tilde{h} ), \nonumber \\   
  \Gamma_{13} &=& 0, \nonumber \\
  \Gamma_{22} &=& \frac{1}{4} \left[ \frac{1}{|\mu_+|}
 \left( \frac{\partial |\mu_+|}{\partial y} \right)^2 +
 \frac{1}{|\mu_-|} \left( \frac{\partial |\mu_-|}{\partial y}
 \right)^2 \right] ~( \tilde{h} | \tilde{h} ) + 
 \left( 2 \pi \frac{\partial \Delta t_d}{\partial y} \right)^2
 \left| \mu_- \right| ~( f \tilde{h} | f \tilde{h} ), \nonumber \\   
  \Gamma_{23} &=& \frac{1}{2} \frac{\partial}{\partial y} \left( |\mu_+|
 + |\mu_-| \right) ~( \tilde{h} | \tilde{h} ), \nonumber \\   
  \Gamma_{33} &=& \left( |\mu_+| + |\mu_-| \right) 
 ~( \tilde{h} | \tilde{h} ),
\label{a2}
\eeqa
and $\Gamma_{ji}=\Gamma_{ij}$. $( \tilde{h} | \tilde{h} )$ and 
 $( f \tilde{h} | f \tilde{h} )$ in the above equation (\ref{a2}) are,
\beqa
  && ( \tilde{h} | \tilde{h} ) = \left( S/N \right)^2 = 4 \int
 \frac{df}{Sn(f)}~ \left| \tilde{h}(f) \right|^2, \nonumber \\
  && ( f \tilde{h} | f \tilde{h} ) = 4 \int \frac{df}{Sn(f)}~
 \left| f \tilde{h}(f) \right|^2. 
\label{a3}
\eeqa
The $S/N$ is the signal to noise ratio for the unlensed signal $\tilde{h}$.
The estimation errors can be analytically obtained by the inverse of the
 Fisher matrix, $\Delta M_{L z}/M_{L z}=[ (\Gamma^{-1})_{11} ]^{1/2}$ 
 and $\Delta y/y=[ (\Gamma^{-1})_{22} ]^{1/2}/y$.
Using the geometrical optics approximation, $f \Delta t_d \gg 1$, we
 obtain the errors with Eq.(\ref{a2}) and (\ref{a3}) as 
\beqa
  \frac{\Delta M_{L z}}{M_{L z}} &=& \frac{1}{S/N} \times \sqrt{
 \frac{|\mu_+|+|\mu_-|}{|\mu_+ \mu_-|}} 
 \left| 2 \frac{\partial}{\partial y} \ln \Delta t_d \Bigg/
 \frac{\partial}{\partial y} \ln \left| \frac{\mu_+}{\mu_-} 
 \right| \right|,   \nonumber \\
  \frac{\Delta y}{y} &=& \frac{1}{S/N} \times \sqrt{
 \frac{|\mu_+|+|\mu_-|}{|\mu_+ \mu_-|}} 
 \left| 2 \Bigg/ y \frac{\partial}{\partial y} \ln \left|
 \frac{\mu_+}{\mu_-} \right| \right|.
\label{a1} 
\eeqa
The above equations (\ref{a1}) are used for the general lens model
 when the double images form.


\begin{figure}
  \plotone{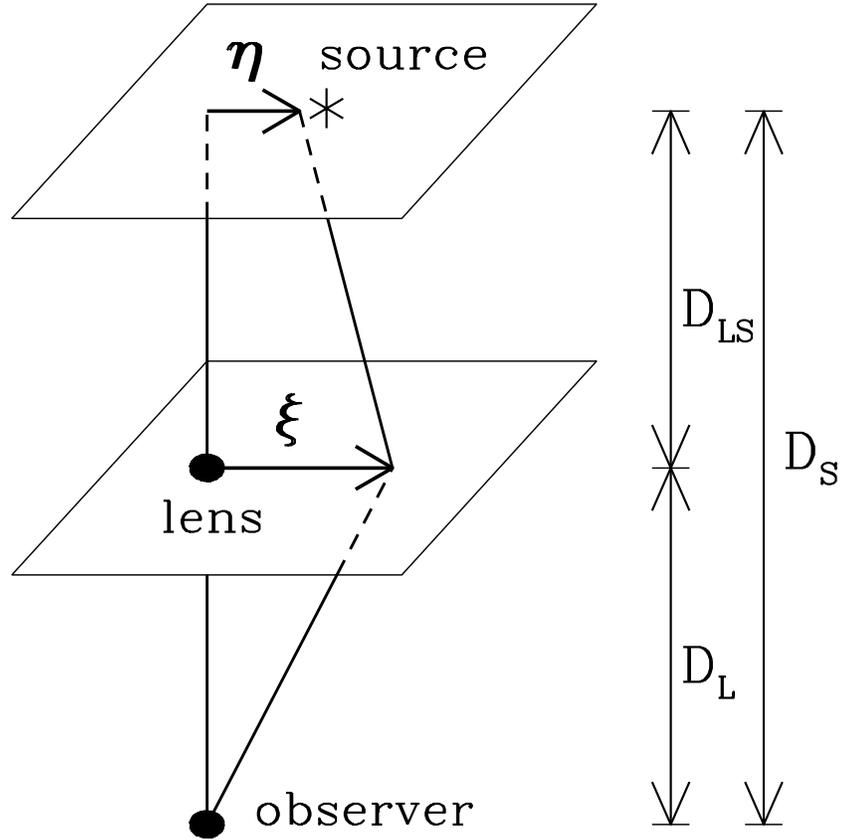}
  \caption{Gravitational lens geometry for the source, the lens and
 the observer.
$D_L, D_S$ and $D_{LS}$ are the distances between them.
$\bfeta$ is a displacement of the source 
 and $\bfxi$ is an impact parameter.
We use the thin lens approximation in which the gravitational waves
 are scattered
 in the thin lens plane. }
\end{figure}

\begin{figure}
  \begin{minipage}{7.5cm}
    \vspace{0.1cm}
    \includegraphics[height=7.5cm,clip]{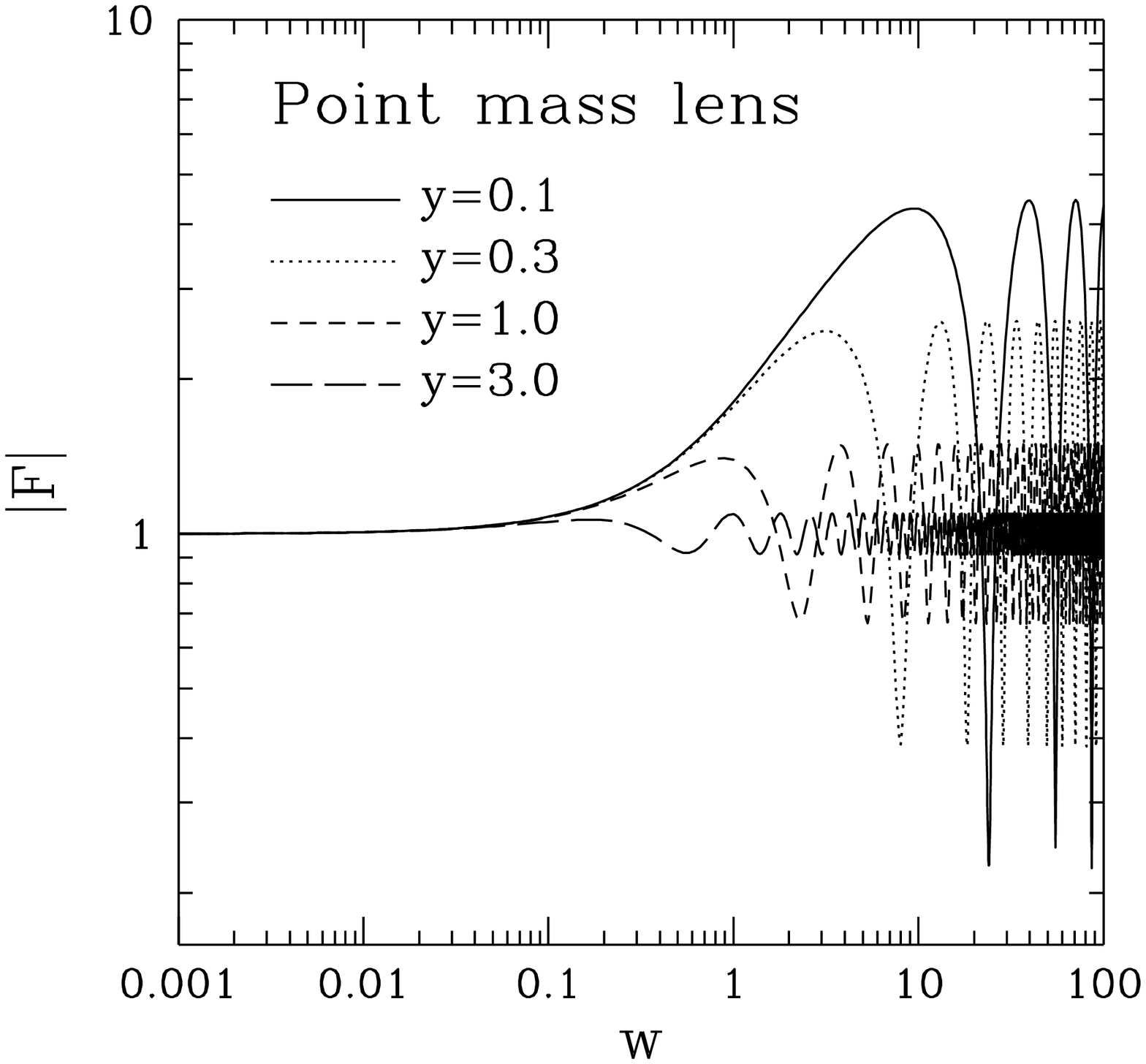} 
    \vspace{0.1cm}
  \end{minipage}
  \begin{minipage}{7.5cm}
    \vspace{0.1cm}
    \includegraphics[height=7.5cm,clip]{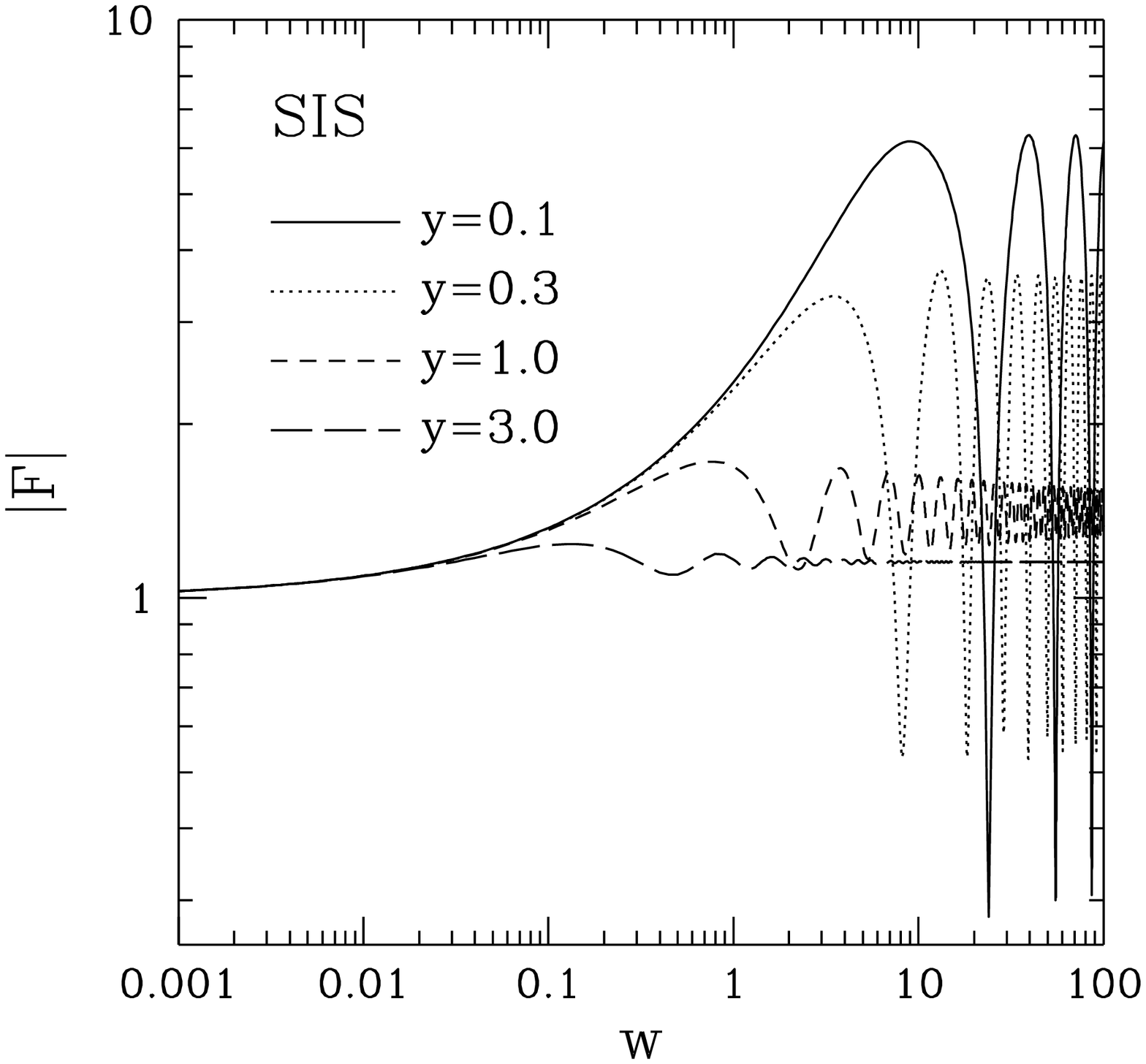} 
    \vspace{0.1cm}
  \end{minipage}
    \\
  \caption{The amplification factor $|F(f)|$ as a function of $w~(=8 \pi
 M_{L z} f)$ with the fixed source position $y=0.1,0.3,1,3$
 for the point mass lens (left panel) and the SIS (right panel).
For $w \lsim 1$, the amplification is very small due to the diffraction
 effect.
For $w \gsim 1$, the oscillatory behavior appears due to the
 interference between the double images.  
We note for the SIS that even if $y \ge 1$ (a single image is formed in the
 geometrical optics limit) the damped oscillatory behavior appears.
}
  \begin{minipage}{7.5cm}
    \vspace{0.1cm}
    \includegraphics[height=7.5cm,clip]{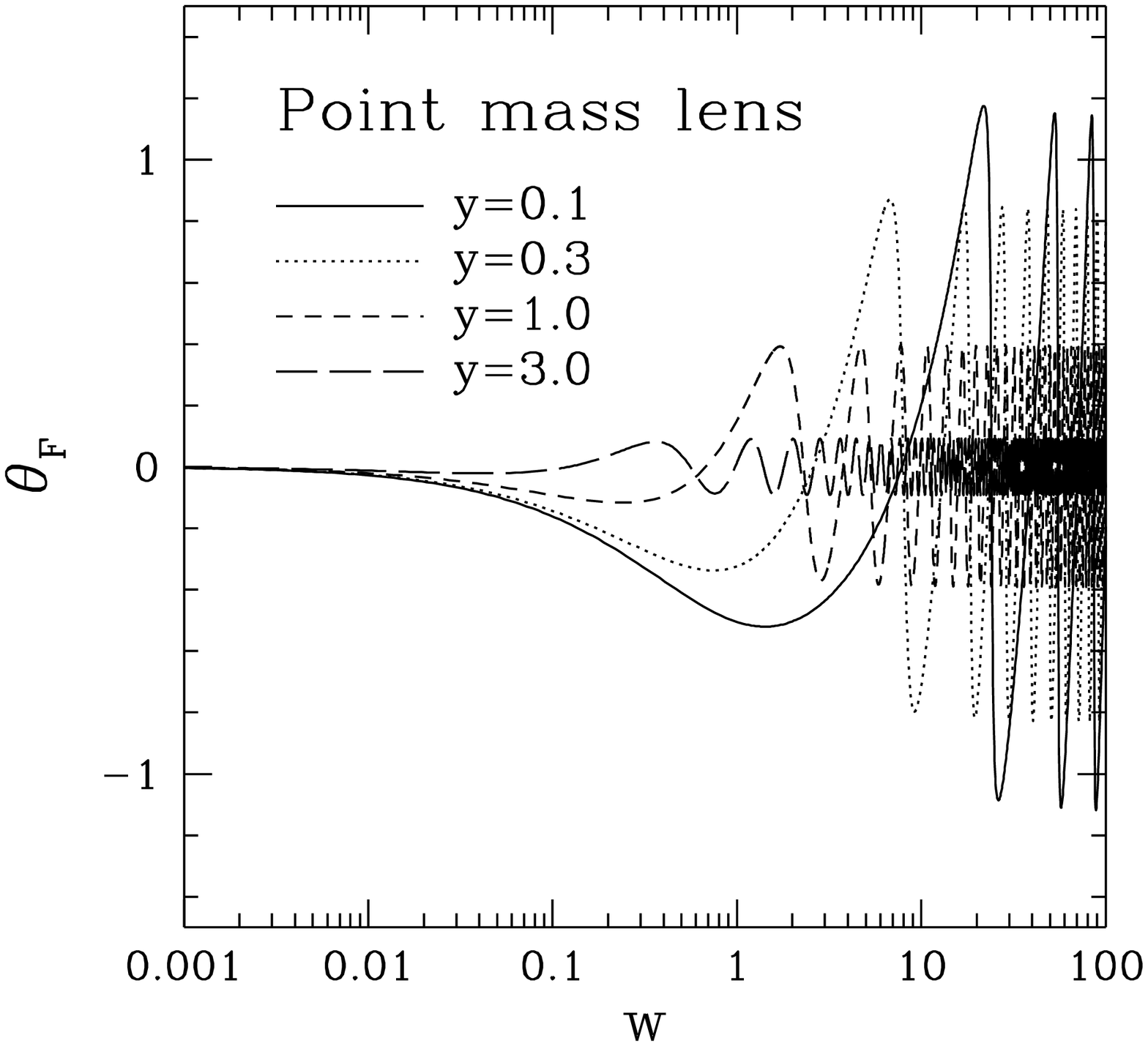} 
    \vspace{0.1cm}
  \end{minipage}
  \begin{minipage}{7.5cm}
    \vspace{0.1cm}
    \includegraphics[height=7.5cm,clip]{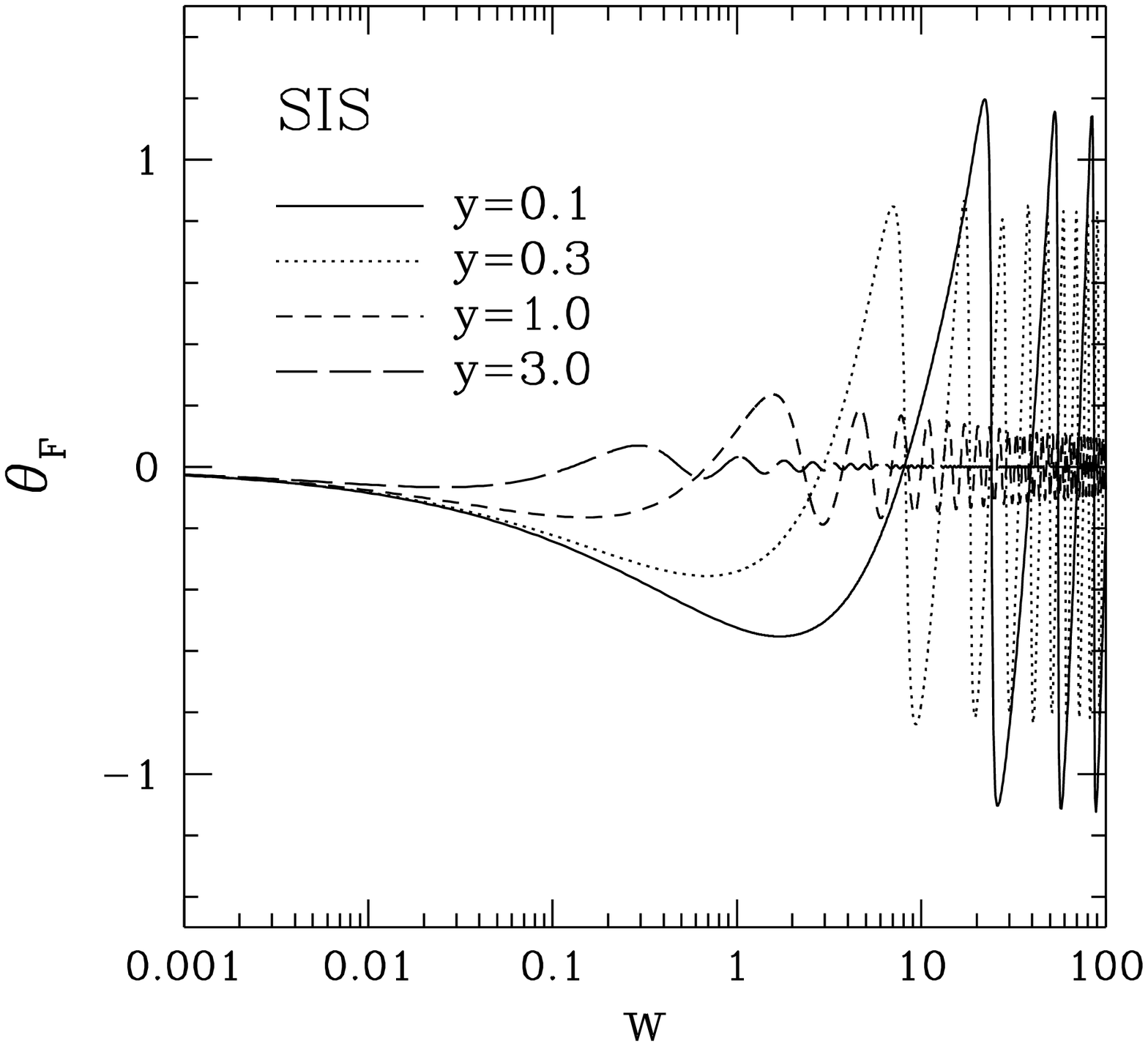} 
    \vspace{0.1cm}
  \end{minipage}
    \\
  \caption{Same as Fig.2, but the phase of the amplification factor
 $\theta_F(f)=-i \ln [F(f)/ |F(f)|]$ as a function of $w~(=8 \pi
 M_{L z} f)$.
}
\end{figure}

\begin{figure}
  \plotone{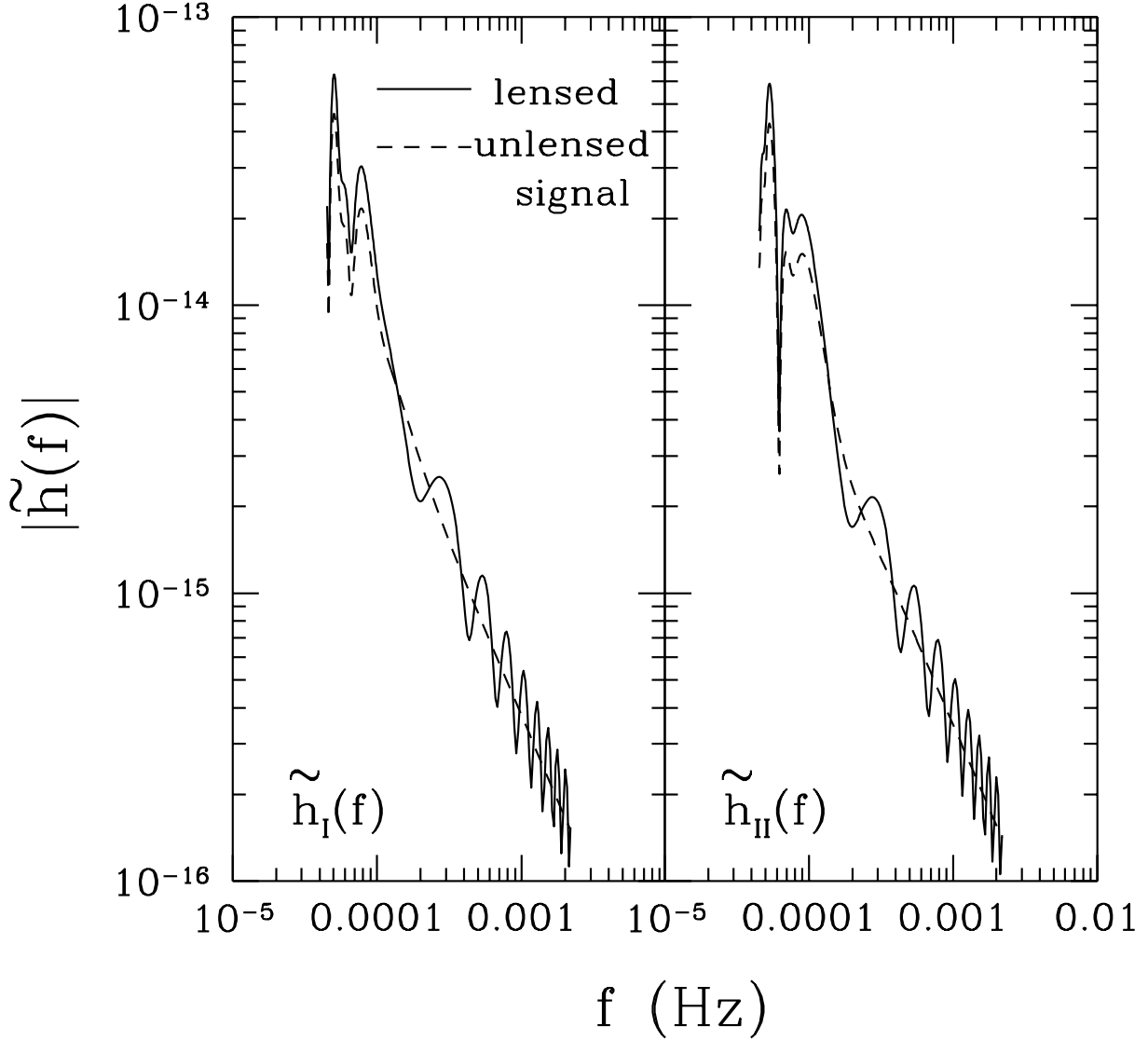}
  \caption{The lensed signals
 $|\tilde{h}^L_{\alpha}(f)|$ $(\alpha=I,II)$
(solid line) and unlensed ones $|\tilde{h}_{\alpha}(f)|$ (dashed line)
 measured by LISA. 
The signals are shown from 1yr before coalescence to 
 ISCO of $r=6 (M_1+M_2)$. 
The redshifted masses of the SMBH binary is $M_{1,2 z}=10^6 M_{\odot}$, 
 the redshifted lens mass is $M_{L z}=10^8 M_{\odot}$ and the source
 position $y=1$. 
The angular parameters are $\cos \bar{\theta}_S=0.3, \bar{\phi}_S=5.0,
 \cos \bar{\theta}_L=0.8, \bar{\phi}_L=2.0$, and the source redshift is
 $z_S=1$ (distance is $H_0 D_S=0.386$). 
The strange behavior for $f \lsim 10^{-4}$ Hz is due to
 the LISA orbital motion, and the difference between the two signals
 is small due to the diffraction.
On the other hand, the oscillatory behavior appears for
 $f \gsim 10^{-4}$ Hz which is 
 determined by the inverse of the lens mass
 $8 \pi M_{L z}$ (see Fig.2). 
This oscillation is due to the interference between the double images. 
} 
\end{figure}

\begin{figure}
  \begin{minipage}[t]{7.5cm}
    \vspace{0.1cm}
    \includegraphics[height=7.5cm,clip]{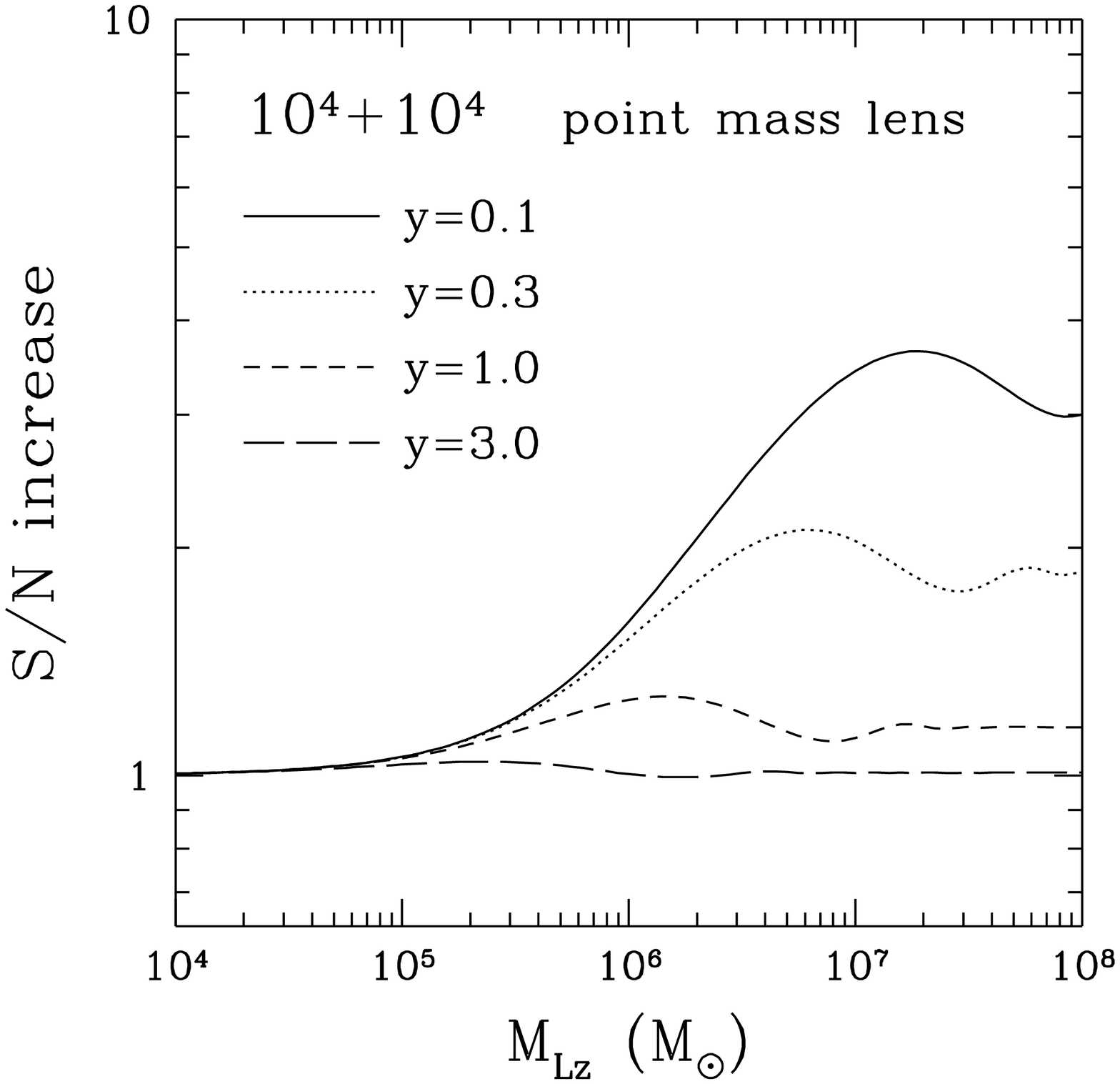} 
    \vspace{0.1cm}
  \end{minipage}
  \begin{minipage}[t]{7.5cm}
    \vspace{0.1cm}
    \includegraphics[height=7.5cm,clip]{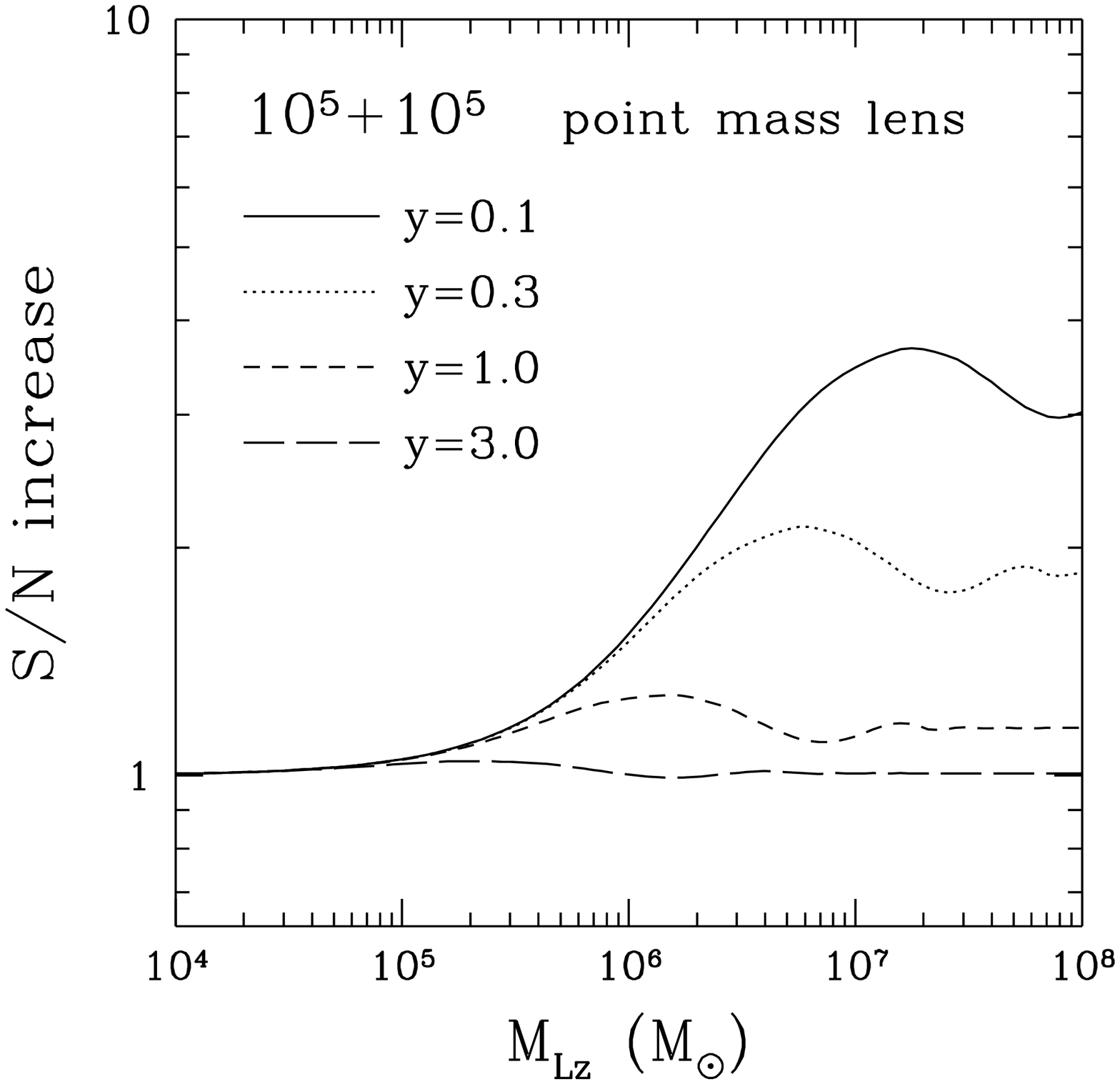} 
    \vspace{0.1cm}
  \end{minipage}
    \\
  \begin{minipage}[t]{7.5cm}
    \vspace{0.1cm}
    \includegraphics[height=7.5cm,clip]{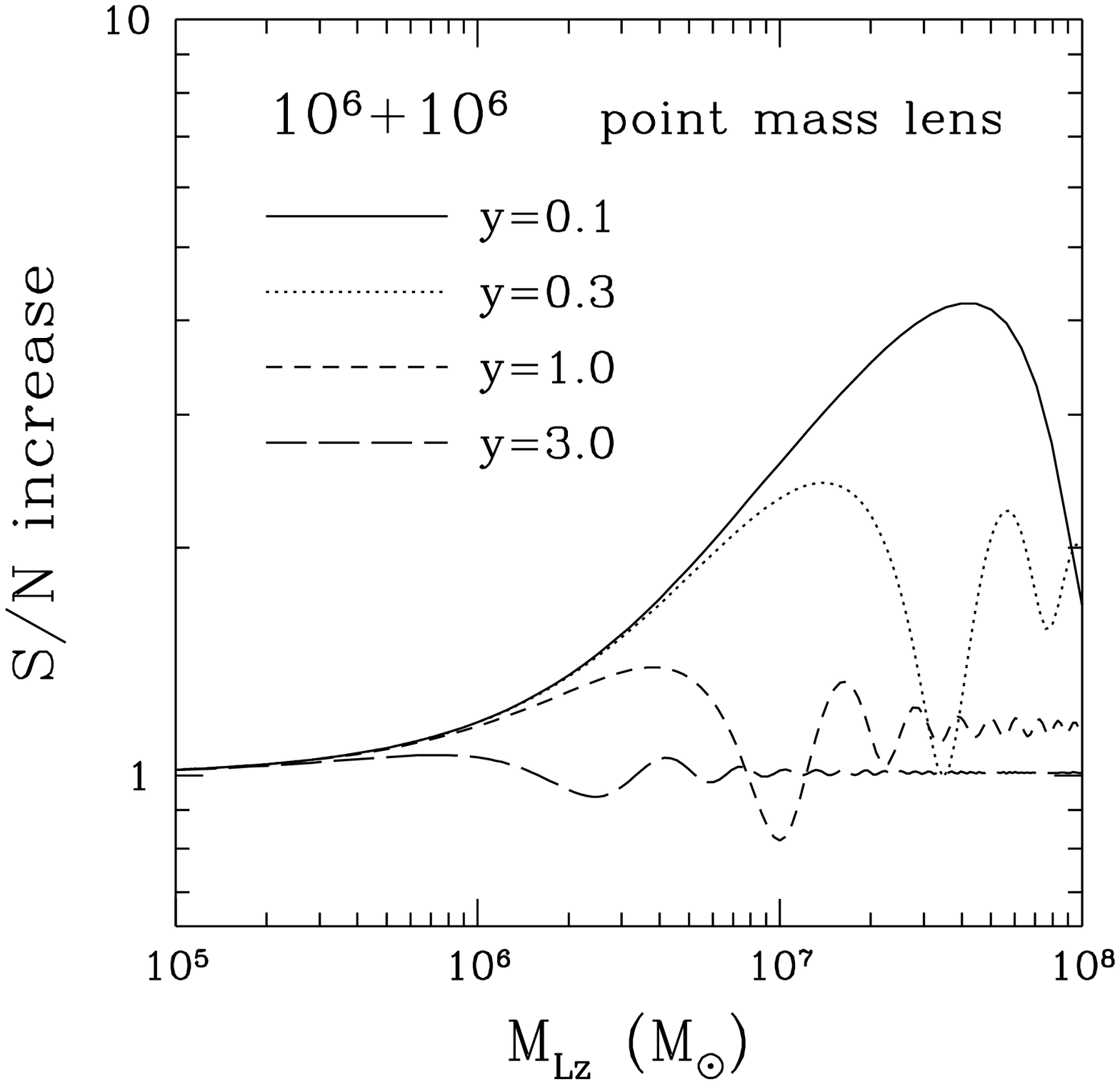} 
    \vspace{0.1cm}
  \end{minipage}
  \begin{minipage}[t]{7.5cm}
    \vspace{0.1cm}
    \includegraphics[height=7.5cm,clip]{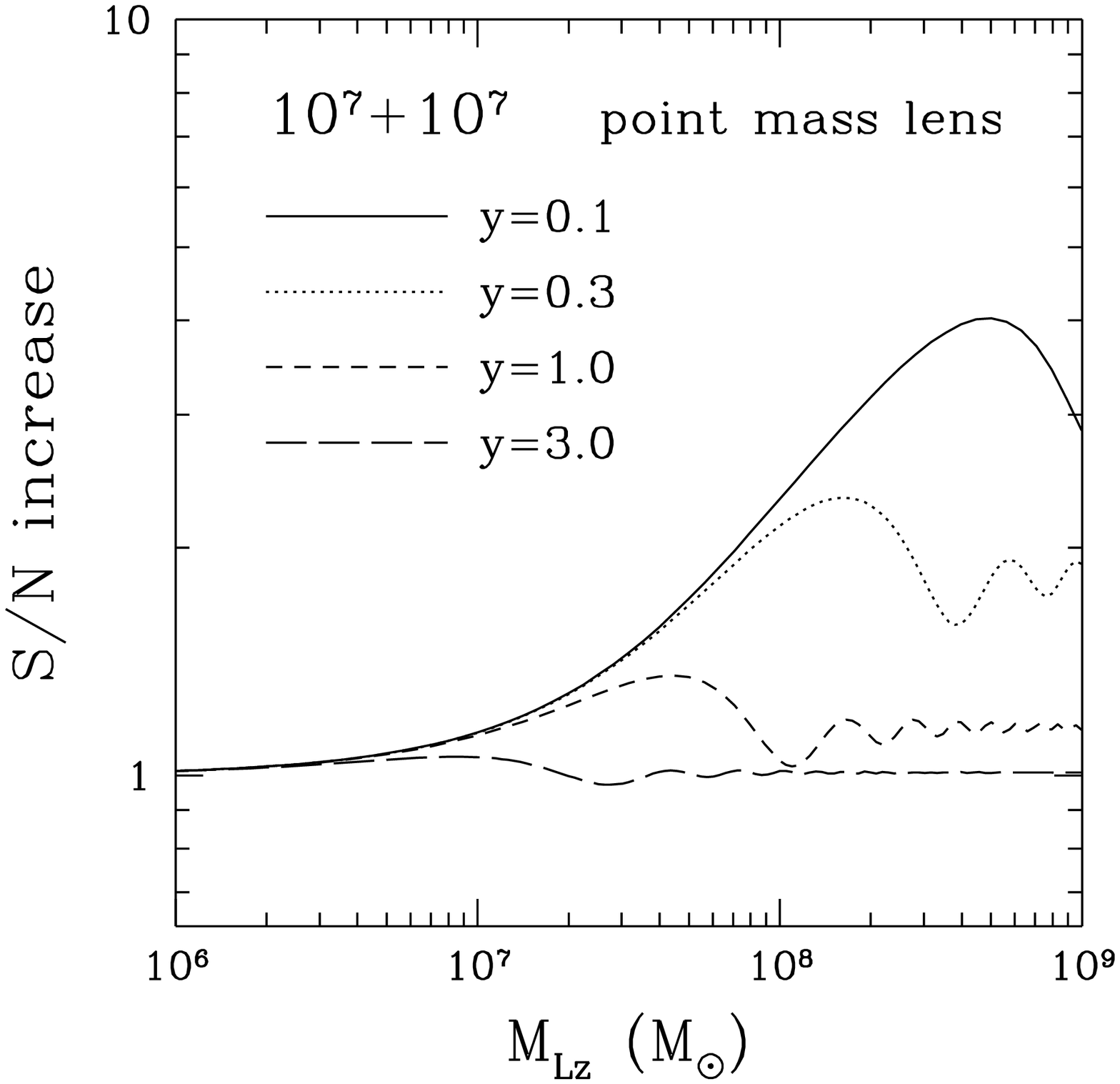} 
    \vspace{0.1cm}
  \end{minipage}
    \\
  \caption{The increasing factor in $S/N$ due to the gravitational 
 lensing by the point mass lens for the various SMBH
 binary masses $M_{1,2 z}=10^4,10^5,10^6,10^7 M_{\odot}$.
The horizontal axis is the redshifted lens mass; the vertical axis is
 the lensed $S/N$ divided by the unlensed $S/N$.
The source position is fixed at $y=0.1, 0.3, 1, 3$. 
For $M_{L z} \lsim 10^6 M_{\odot}$, the
 magnification is very small due to the diffraction effect 
 irrespective of the SMBH binary masses.
For $M_{L z} \gsim 10^7 M_{\odot}$, 
 the damped oscillatory patterns appear due to the interference between
 the two images, and this behavior
 converge in the geometrical optics
 limit, $|\mu|^{1/2}=(y^2+2)^{1/2}/[y^{1/2} (y^2+4)^{1/4}]$.
} 

\end{figure}

\begin{figure}
  \begin{minipage}[t]{7.5cm}
    \vspace{0.1cm}
    \includegraphics[height=7.5cm,clip]{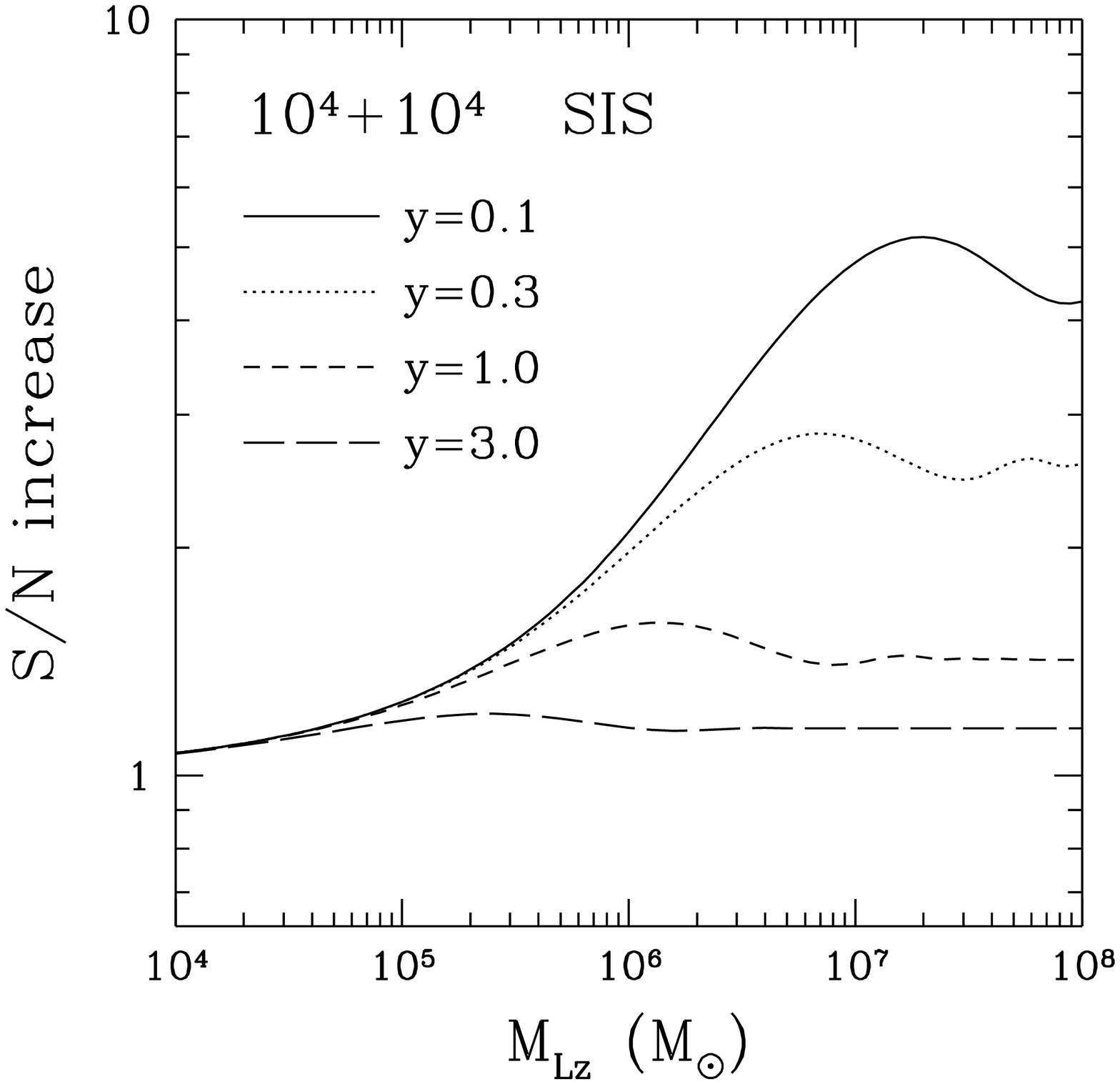} 
    \vspace{0.1cm}
  \end{minipage}
  \begin{minipage}[t]{7.5cm}
    \vspace{0.1cm}
    \includegraphics[height=7.5cm,clip]{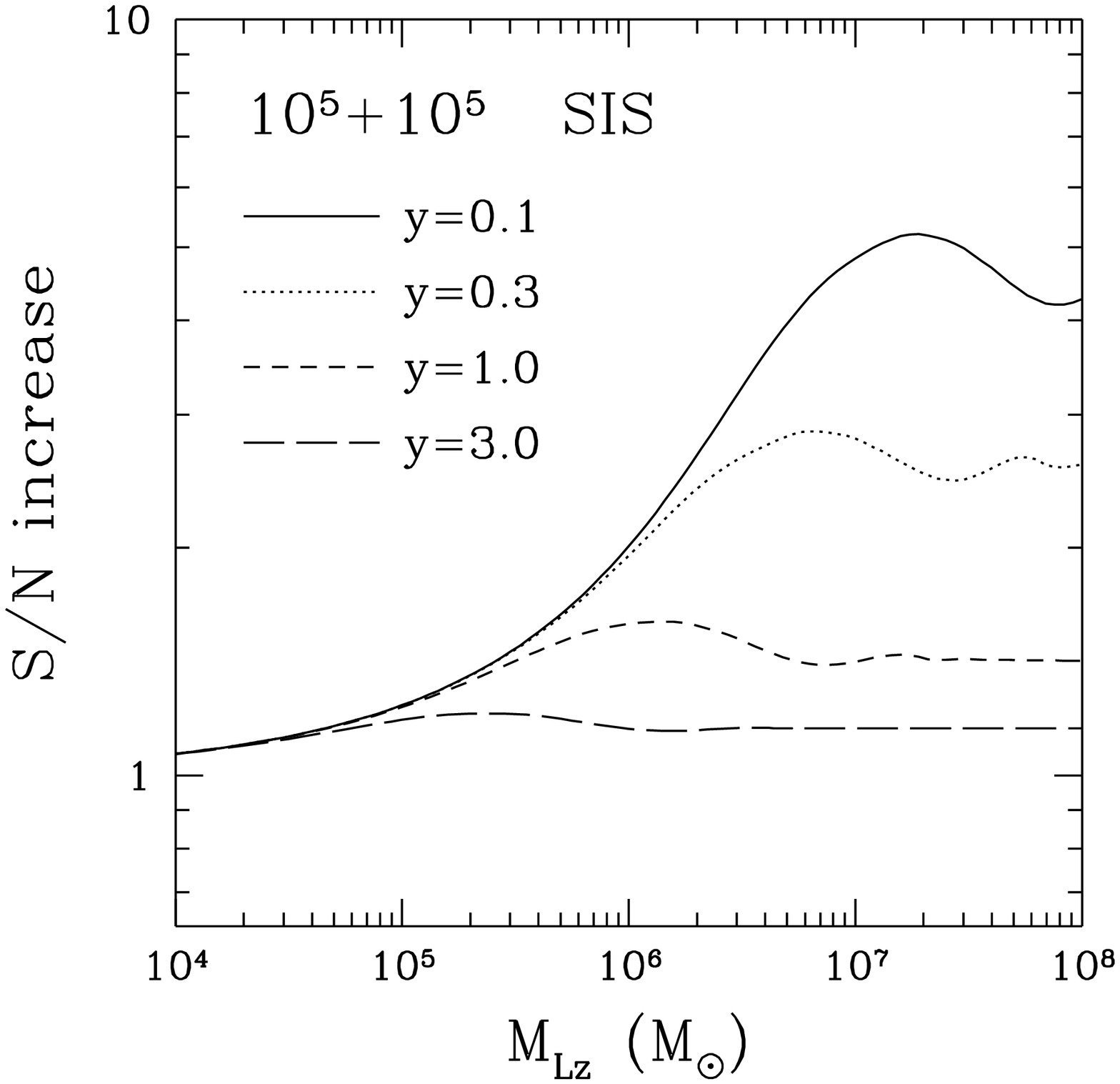} 
    \vspace{0.1cm}
  \end{minipage}
    \\
  \begin{minipage}[t]{7.5cm}
    \vspace{0.1cm}
    \includegraphics[height=7.5cm,clip]{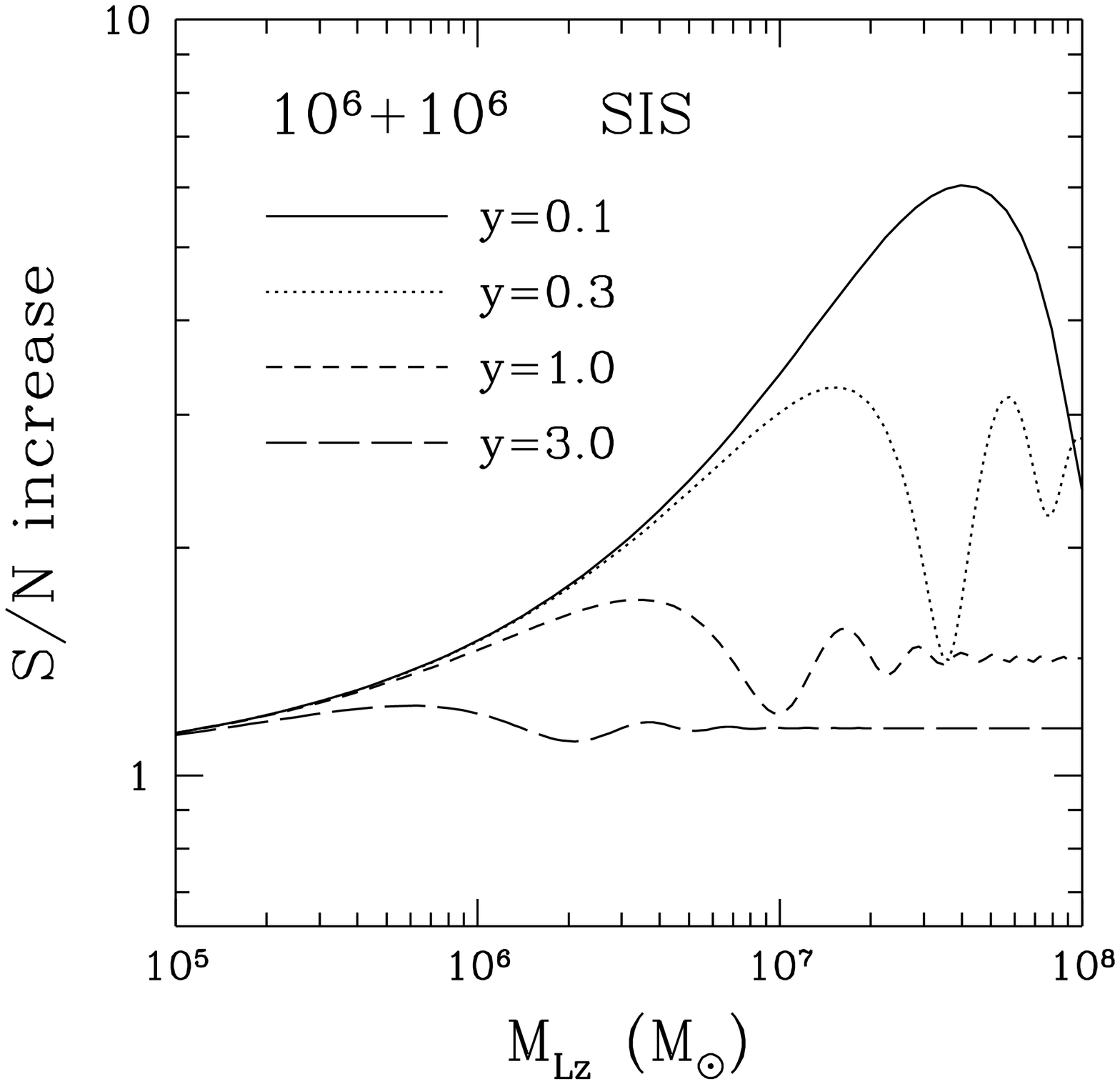} 
    \vspace{0.1cm}
  \end{minipage}
  \begin{minipage}[t]{7.5cm}
    \vspace{0.1cm}
    \includegraphics[height=7.5cm,clip]{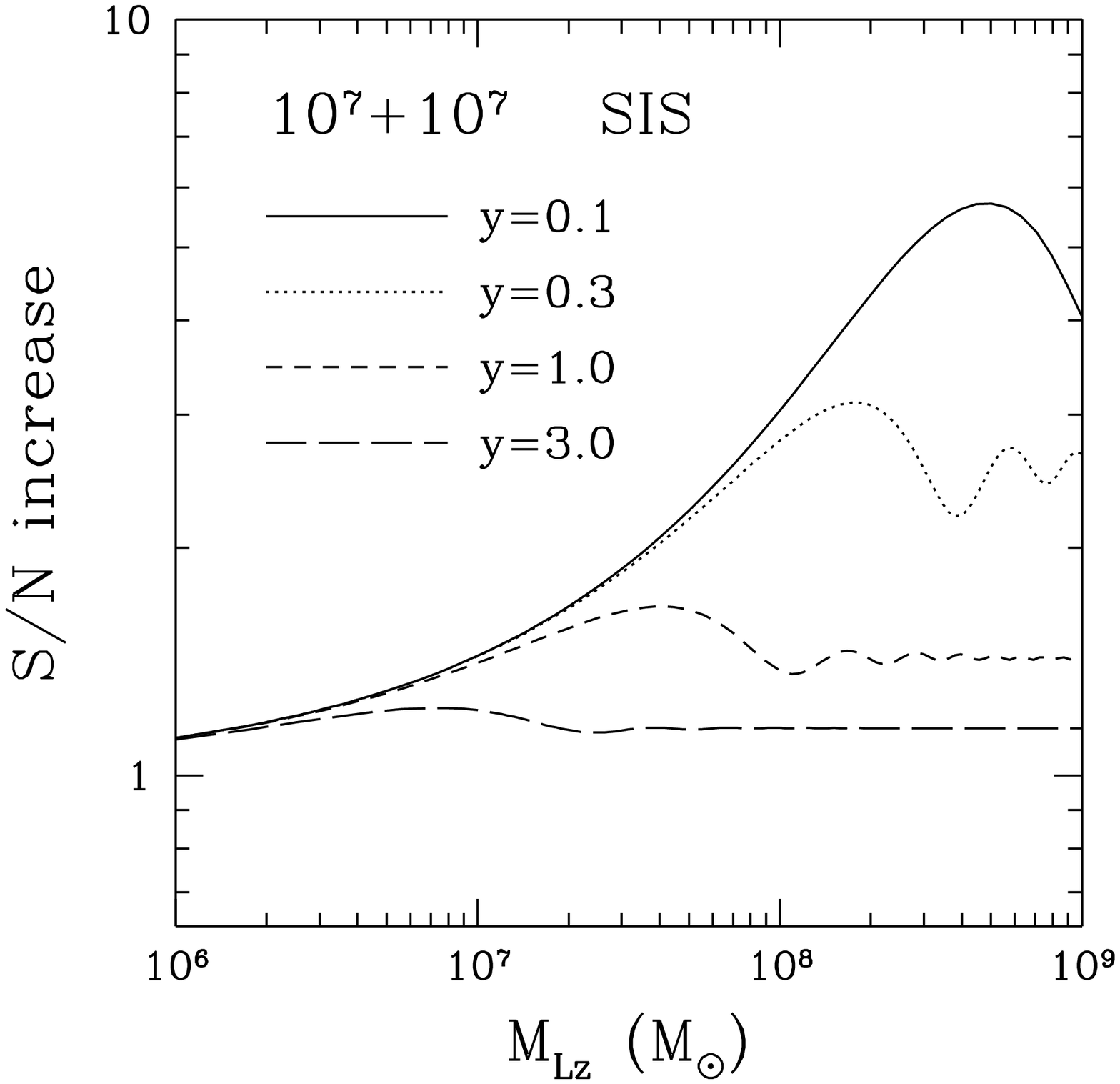} 
    \vspace{0.1cm}
  \end{minipage}
    \\
  \caption{Same as Fig.5, but for the SIS lens model.
For $M_{L z} \gsim 10^7 M_{\odot}$, 
 the results converge in the geometrical optics
 limit, $|\mu|^{1/2}=(2/y)^{1/2}$ for $y \leq 1$ and
 $|\mu|^{1/2}=(1+1/y)^{1/2}$ for $y \geq 1$.
} 
\end{figure}

\begin{figure}
  \begin{minipage}[t]{7.5cm}
    \vspace{0.1cm}
    \includegraphics[height=7.5cm,clip]{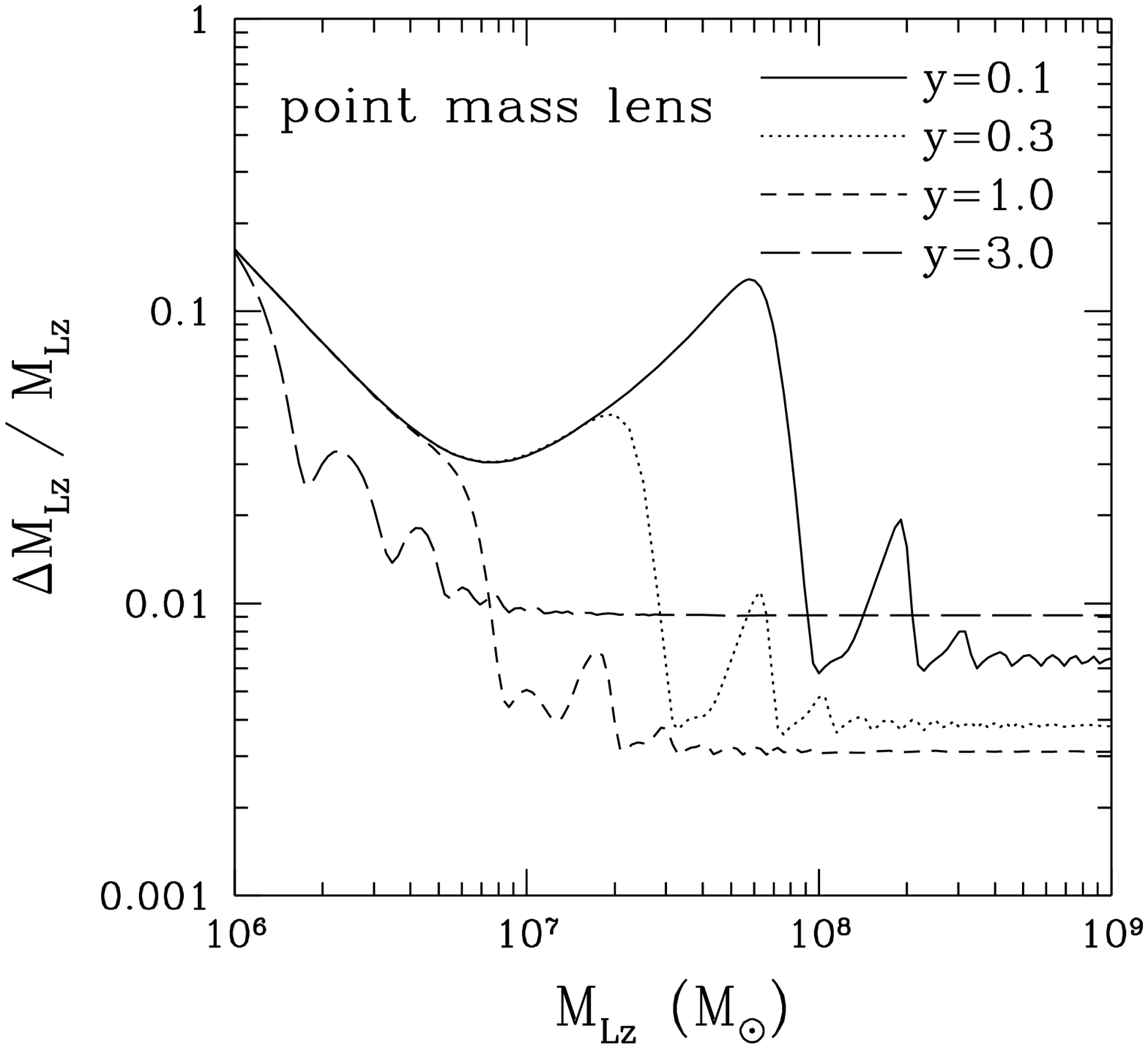} 
    \vspace{0.1cm}
  \end{minipage}
  \begin{minipage}[t]{7.5cm}
    \vspace{0.1cm}
    \includegraphics[height=7.5cm,clip]{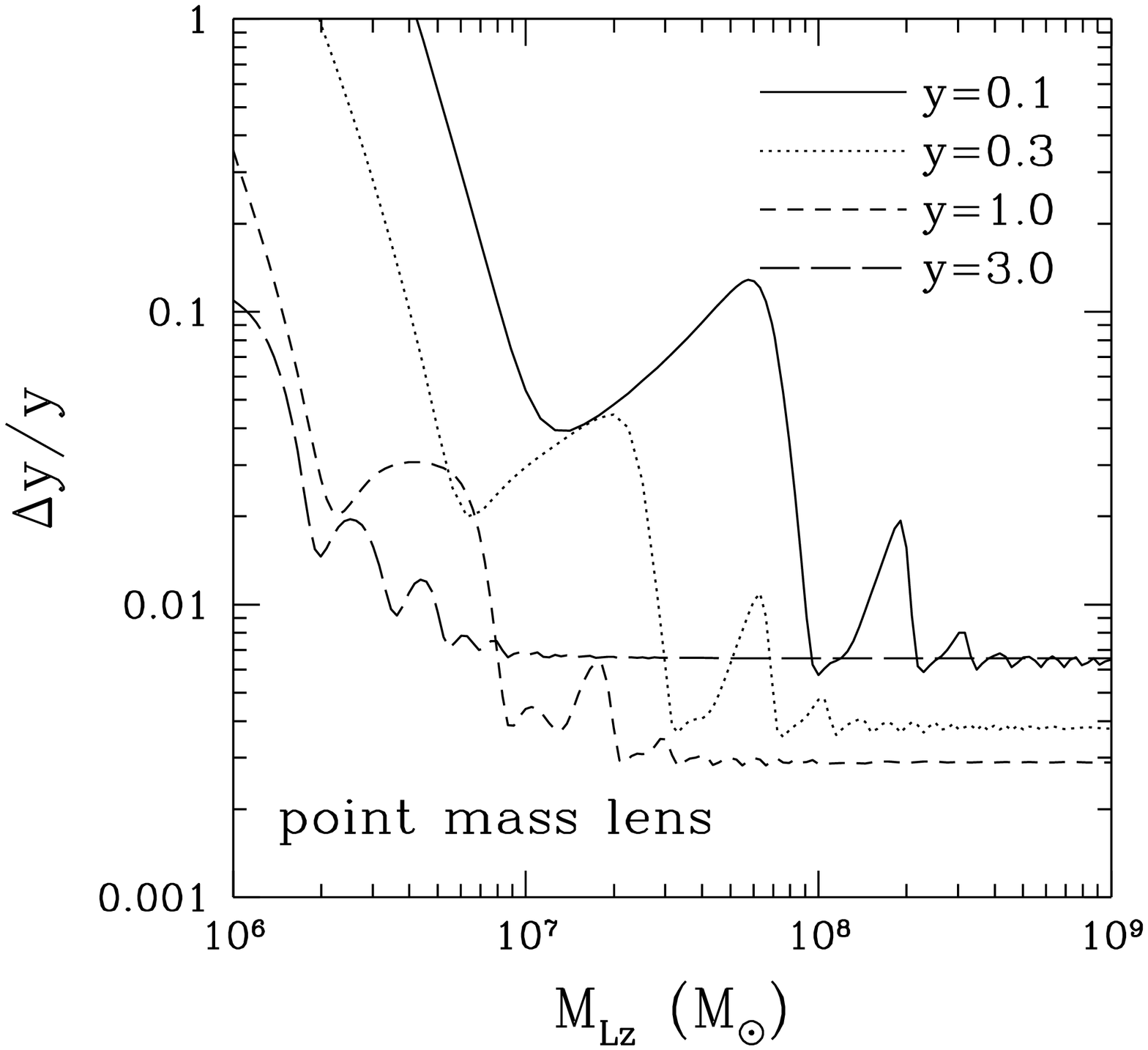} 
    \vspace{0.1cm}
  \end{minipage}
    \\
  \caption{The estimation errors for the redshifted lens mass
 $\Delta M_{L z}$ (left panel) and the source position $\Delta y$
 (right panel) for the point mass lens.
The results are presented for the SMBH binary of masses
 $10^6 + 10^6 M_{\odot}$ at $z_S=1$. 
The errors are normalized by $S/N=10^3$
 and simply scale as $(S/N)^{-1}$.
For $M_{L z} \lsim 10^{7} M_{\odot}$ the errors are
 relatively large, since the effect of lensing is
 very small due to the diffraction.
For $M_{L z} \gsim 10^8 M_{\odot}$ the geometrical optics approximation
 is valid, and errors converge to constants.
}
  \begin{minipage}[t]{7.5cm}
    \vspace{0.1cm}
    \includegraphics[height=7.5cm,clip]{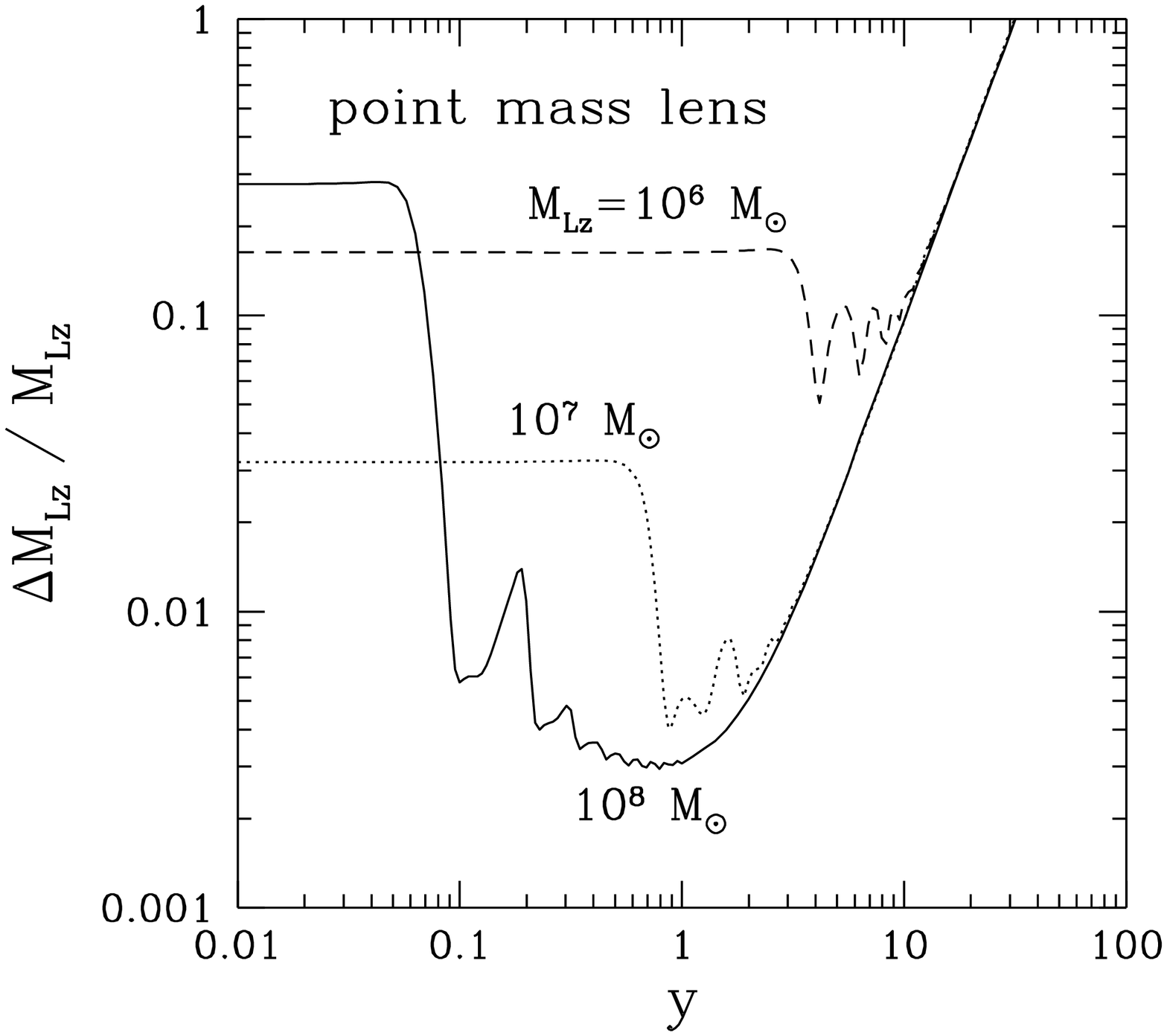} 
    \vspace{0.1cm}
  \end{minipage}
  \begin{minipage}[t]{7.5cm}
    \vspace{0.1cm}
    \includegraphics[height=7.5cm,clip]{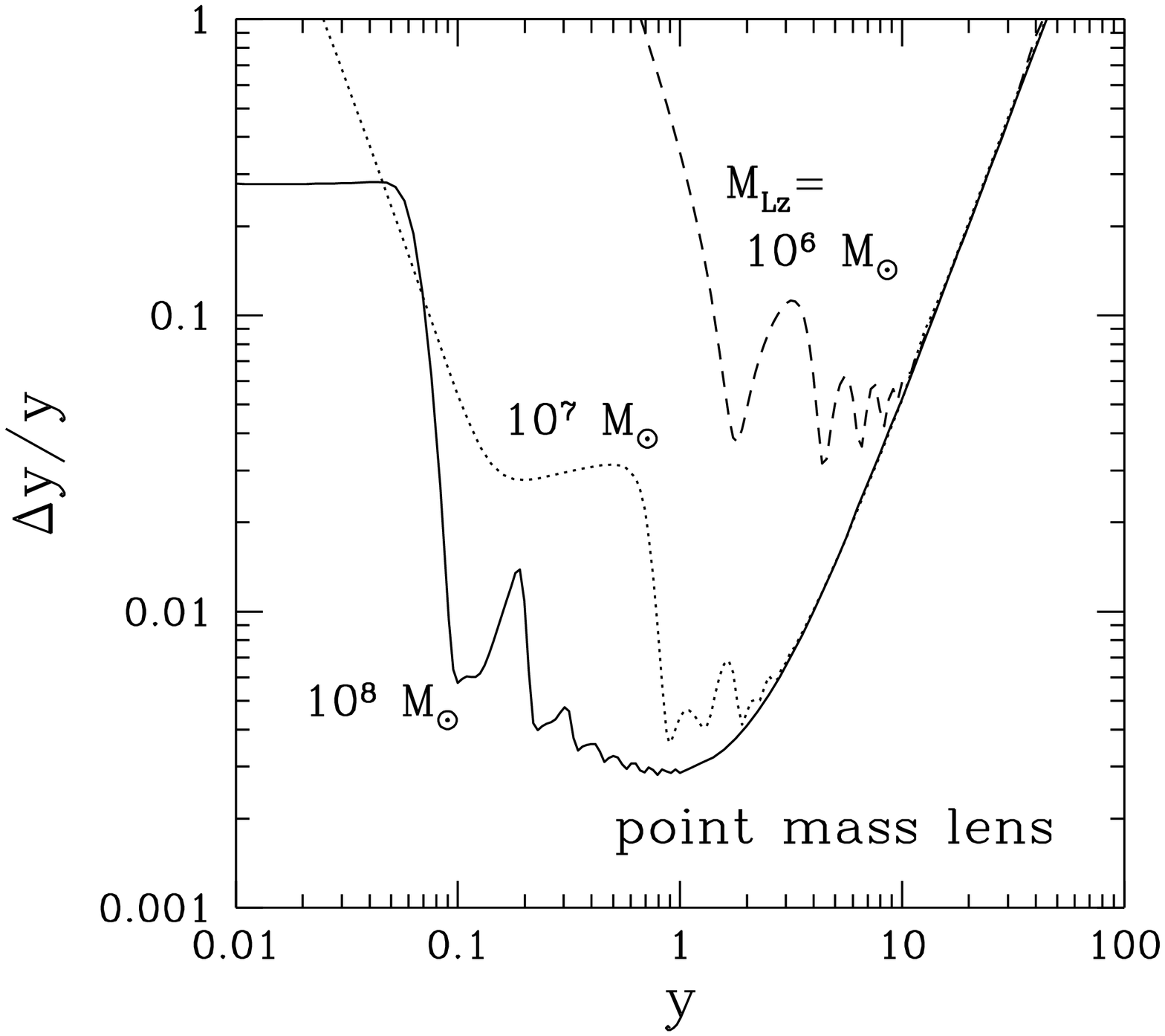} 
    \vspace{0.1cm}
  \end{minipage}
    \\
  \caption{Same as Fig.7, but as a function of $y$.
We note that even for $y \gsim 10$ we can extract the lens
 information.
Then the lensing cross section ($\propto y^2$) increases an order
 of magnitude larger than 
 that for the usual strong lensing of light ($y=1$).
}
\end{figure}

\begin{figure}
  \begin{minipage}[t]{7.5cm}
    \vspace{0.1cm}
    \includegraphics[height=7.5cm,clip]{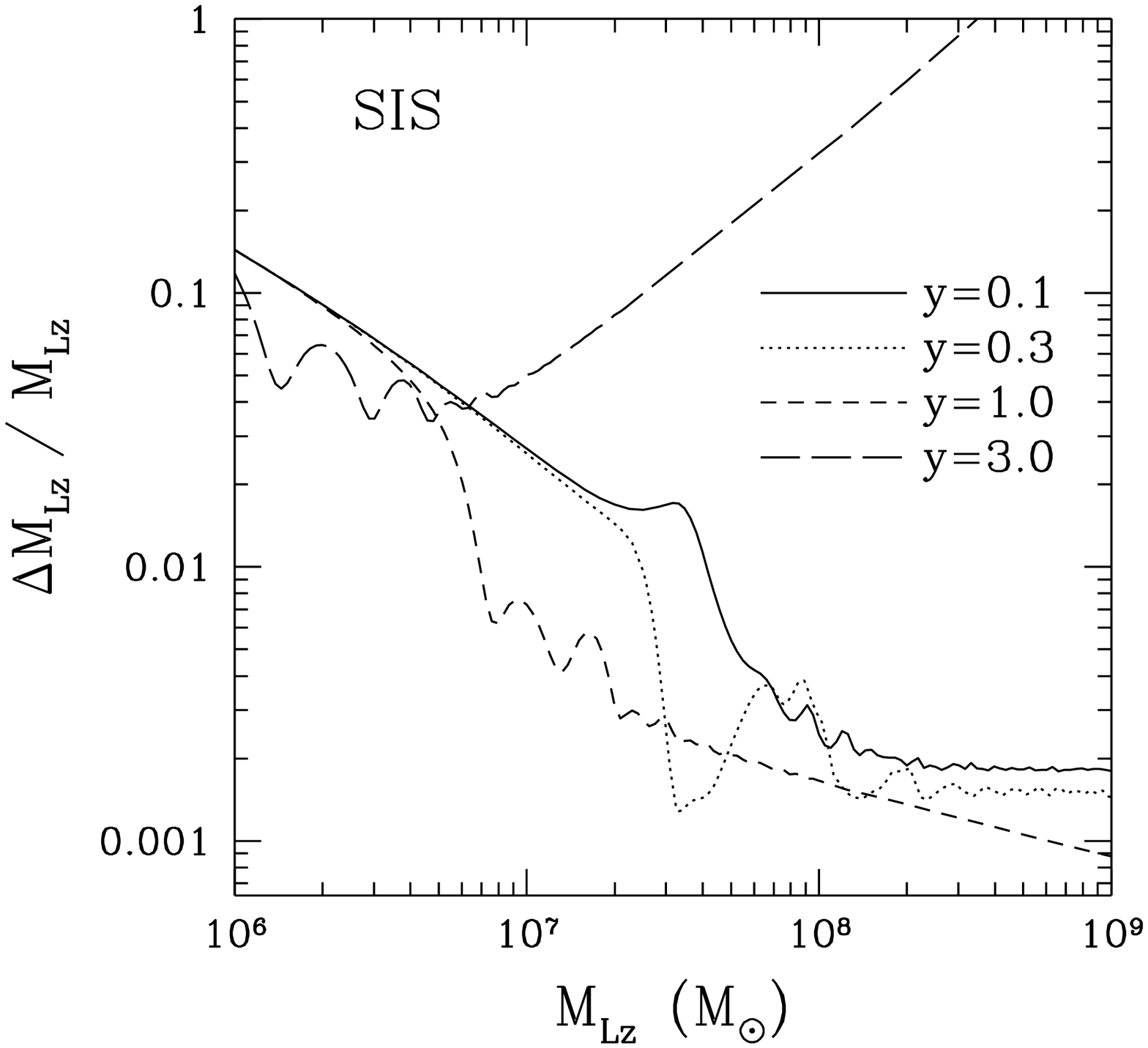} 
    \vspace{0.1cm}
  \end{minipage}
  \begin{minipage}[t]{7.5cm}
    \vspace{0.1cm}
    \includegraphics[height=7.5cm,clip]{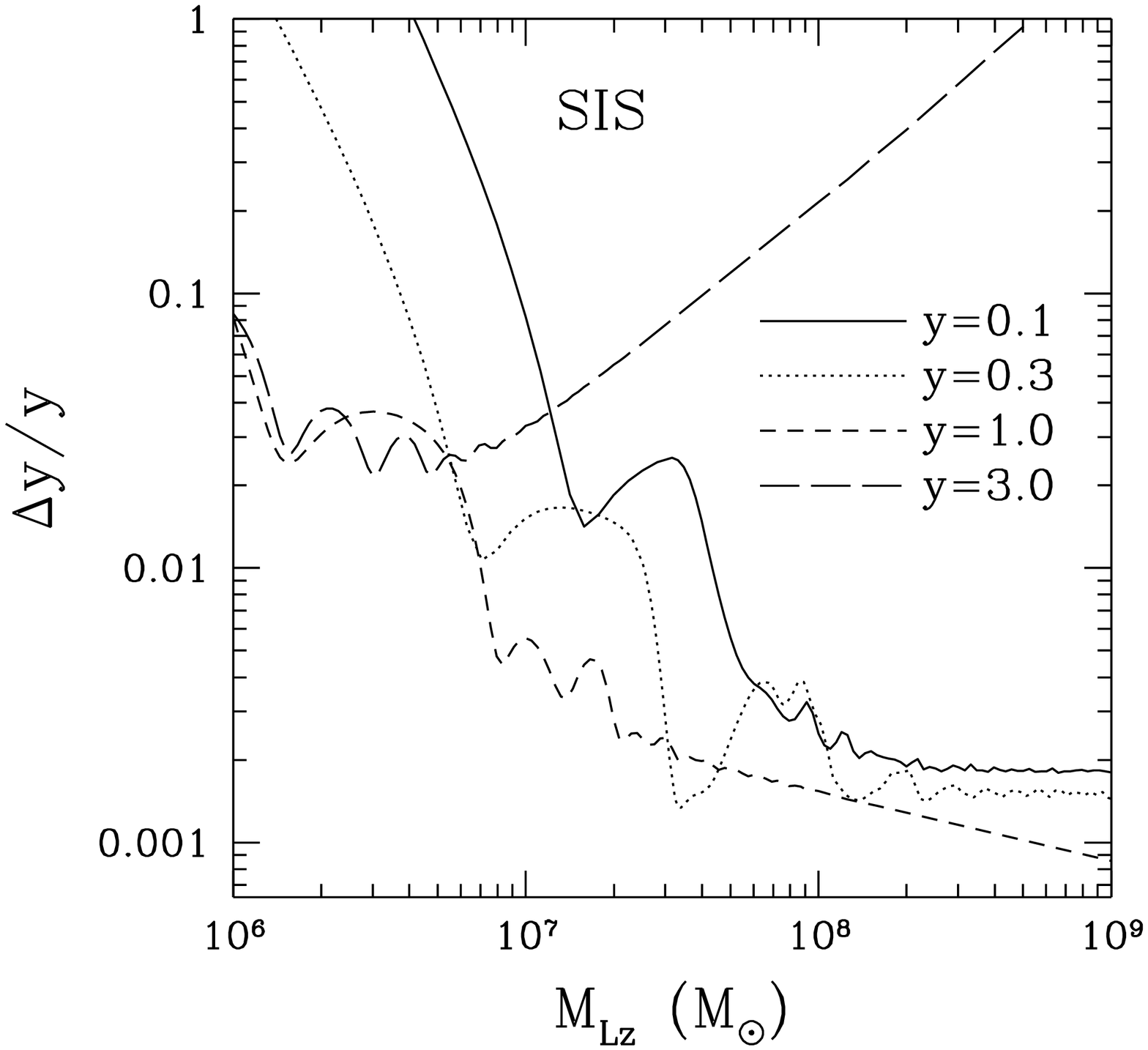} 
    \vspace{0.1cm}
  \end{minipage}
    \\
  \caption{The estimation errors for the redshifted lens mass
 $\Delta M_{L z}$ (left) and the source position $\Delta y$ (right)
 for the SIS model.
The results are presented for the SMBH binary of masses
 $10^6 + 10^6 M_{\odot}$ at $z_S=1$.
The errors are normalized by $S/N=10^3$
 and simply scale as $(S/N)^{-1}$.
Even for $y=3$ (a single image is formed in the geometrical optics limit),
 the lens parameters can 
 be extracted at $M_{L z} \sim 10^{6} - 10^{8} M_{\odot}$
 due to the wave effects.  
}
  \begin{minipage}[t]{7.5cm}
    \vspace{0.1cm}
    \includegraphics[height=7.5cm,clip]{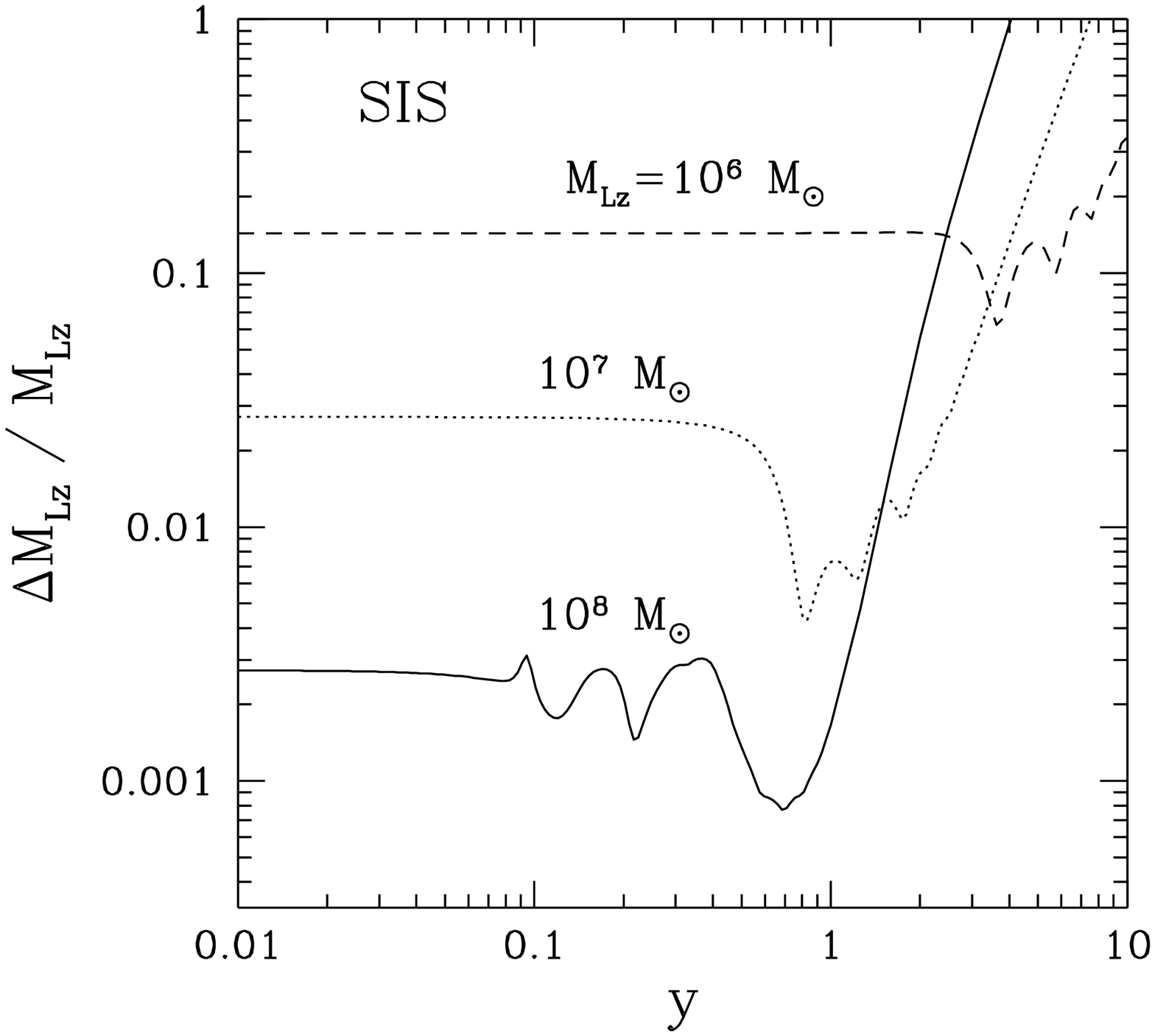} 
    \vspace{0.1cm}
  \end{minipage}
  \begin{minipage}[t]{7.5cm}
    \vspace{0.1cm}
    \includegraphics[height=7.5cm,clip]{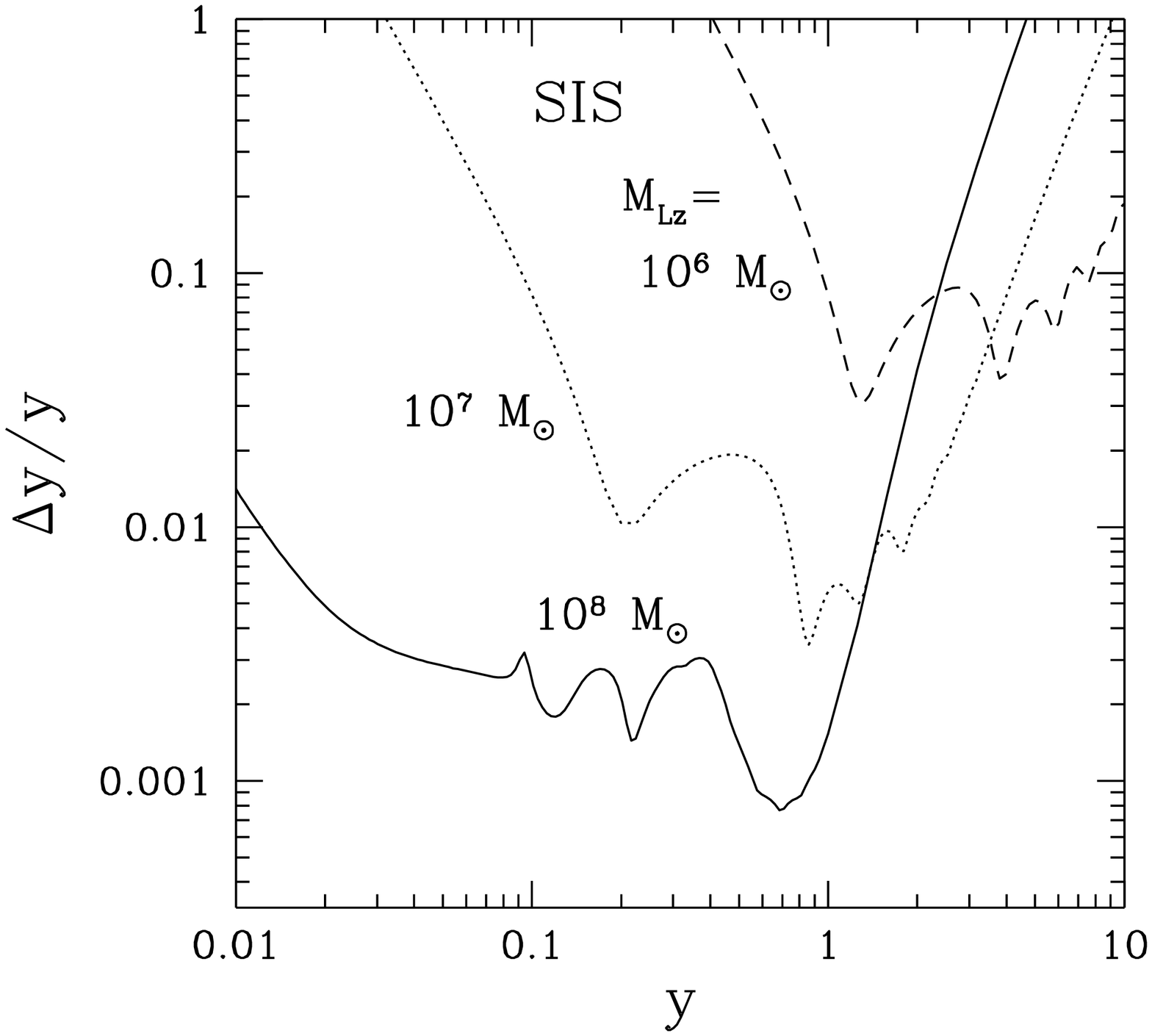} 
    \vspace{0.1cm}
  \end{minipage}
    \\
  \caption{Same as Fig.9, but as a function of $y$.
We note that even for $y > 1$ we can determine the lens
 parameters.
Then the lensing cross section ($\propto y^2$) becomes
 larger than that in the geometrical optics approximation ($y=1$).
}
\end{figure}

\newpage

\begin{table} 
  \begin{center}
  \setlength{\tabcolsep}{10pt}
  \renewcommand{\arraystretch}{1.1}
  \begin{tabular}{ccccc} \hline\hline
   Binary Masses~($M_{\odot}$) & ~$z_S=1$~ & ~$z_S=3$~ & ~$z_S=5$~ &
 ~$z_S=10$ \\ \hline
   $10^7 + 10^7$~ & ~1038~ & ~270~ & ~147~ & ~66~ \\    
   $10^7 + 10^6$~ & ~519~ & ~135~ & ~74~ & ~33~ \\ 
   $10^7 + 10^5$~ & ~175~ & ~46~ & ~25~ & ~11~ \\ 
   $10^7 + 10^4$~ & ~52~ & ~14~ & ~7~ & ~3~ \\
   $10^6 + 10^6$~ & ~2575~ & ~669~ & ~365~ & ~164~ \\    
   $10^6 + 10^5$~ & ~1517~ & ~394~ & ~215~ & ~97~ \\ 
   $10^6 + 10^4$~ & ~508~ & ~132~ & ~72~ & ~32~ \\ 
   $10^5 + 10^5$~ & ~877~ & ~228~ & ~124~ & ~56~ \\
   $10^5 + 10^4$~ & ~310~ & ~81~ & ~44~ & ~20~ \\ 
   $10^4 + 10^4$~ & ~132~ & ~34~ & ~19~ & ~8~ 
  \\ \hline\hline
  \end{tabular}
  \end{center} 
\caption{
The signal to noise ratio (S/N) for the various binary masses
 $10^4-10^7 M_{\odot}$ with redshift $z_S=1,3,5,10$.
We assume 1 yr observation of in-spiral phase
 before final merging.
}
\label{t2}
\end{table}

\newpage

\begin{table}
  \begin{center}
  \setlength{\tabcolsep}{10pt}
  \renewcommand{\arraystretch}{1.1}
  \begin{tabular}{ccccc} \hline\hline
   Lens Model~ & ~$z_S=1$~ & ~$z_S=3$~ & ~$z_S=5$~ & ~$z_S=10$
 \\ \hline
   Point mass lens~ & ~$< 0.21$~ & ~$< 1.1$~ & ~$< 2.0$~ & ~$< 3.9$~ \\    
   SIS~ & ~$7.2 \times 10^{-5}$ ~ & ~$8.1 \times 10^{-4} $~
 & ~$2.0 \times 10^{-3}$~ & ~$4.7 \times 10^{-3} $~
  \\ \hline\hline
  \end{tabular}
  \end{center} 
\caption{
The lensing probability by the lens mass in the range
 $10^6-10^9 M_{\odot}$ with the source redshift $z_S=1,3,5,10$.
For the point mass lens, we give the upper limit which is determined by the
 observational constraint on the abundance of the compact objects.
When the lensing probability is more than one, the lensing
 occurs some times.
For the SIS, CDM halos are assumed to be lenses.
The presented values are for the case of $S/N=10^3$, and hence the
 results are somewhat overestimated for the binaries of $S/N < 10^3$
 in Table.1. 
If the expected rate of merging SMBHs is $\sim 300$ per
 year (Wyithe \& Loeb 2002), 
 then the lensing events will be detected $1$ event per year.
}
\label{t1}
\end{table}

\end{document}